\documentclass[a4paper, twocolumn, 10pt]{article}
\usepackage[english]{babel}

\usepackage[top=2cm, bottom=2cm, left=2cm, right=2cm]{geometry}
\usepackage{braket}
\usepackage{caption}
\usepackage{subcaption}
\usepackage{qcircuit}
\usepackage{xcolor}
\usepackage{verbatim}
\usepackage{adjustbox}

\usepackage{amsmath}
\usepackage{amssymb}
\usepackage{graphicx}
\usepackage{dsfont}
\usepackage{xfrac}
\usepackage{authblk}
\usepackage{pgfplots}
\usepgfplotslibrary{groupplots}
\pgfplotsset{compat=1.18}
\usetikzlibrary{matrix}
\usetikzlibrary{calc}
\usetikzlibrary{plotmarks}

\definecolor{gnomeblue}{RGB}{52,101,164}
\definecolor{gnomeorange}{RGB}{244,156,0}
\definecolor{gnomegreen}{RGB}{126,200,80}

\definecolor{TangoBlue}{HTML}{729FCF}
\definecolor{TangoOrange}{HTML}{FCAF3E}
\definecolor{TangoGreen}{HTML}{8AE234}
\definecolor{TangoPurple}{HTML}{AD7FA8}
\definecolor{TangoRed}{HTML}{EF2929}
\definecolor{TangoButter}{HTML}{FCE94F}
\definecolor{TangoMagenta}{HTML}{CC3399}
\definecolor{TangoCyan}{HTML}{00AEEF}

\definecolor{Purplish}{RGB}{128,0,128}  
\definecolor{Pinkish}{RGB}{255,105,180}   
\definecolor{MyBlue}{RGB}{180,200,255}
\definecolor{MyRed}{RGB}{255,170,170}

\usepackage[normalem]{ulem}
\usepackage[colorlinks=true, allcolors=blue, pagebackref]{hyperref}

\title{Bayesian Quantum Orthogonal Neural Networks for Anomaly Detection}

\author[1, 2, *]{Natansh Mathur}
\author[1]{Brian Coyle}
\author[1]{Nishant Jain}
\author[1, 3]{Snehal Raj}
\author[4]{Akshat Tandon}
\author[4]{Jasper Simon Krauser}
\author[4]{Rainer Stoessel}
\affil[1]{QC Ware, Palo Alto, USA and Paris, France.}
\affil[2]{IRIF, CNRS - Universit\'e Paris Cit\'e, France.}
\affil[3]{LIP6, CNRS - Sorbonne Universit\'e, Paris, France.}
\affil[4]{Central Research and Technology, Airbus Group, Germany.}
\affil[*]{Corresponding author, natansh.mathur@qcware.com}

\date{}

\begin{document}
\maketitle

\begin{abstract}
Identification of defects or anomalies in 3D objects is a crucial task to ensure correct functionality. In this work, we combine Bayesian learning with recent developments in quantum and quantum-inspired machine learning, specifically orthogonal neural networks, to tackle this anomaly detection problem for an industrially relevant use case. Bayesian learning enables uncertainty quantification of predictions, while orthogonality in weight matrices enables smooth training. We develop orthogonal (quantum) versions of 3D convolutional neural networks and show that these models can successfully detect anomalies in 3D objects. To test the feasibility of incorporating quantum computers into a quantum-enhanced anomaly detection pipeline, we perform hardware experiments with our models on IBM's 127-qubit Brisbane device, testing the effect of noise and limited measurement shots.
\end{abstract}

\section{Introduction}

In recent years, the intersection of quantum computing and machine learning has fostered the development of novel computational paradigms that push the boundaries of traditional algorithms \cite{qmarl_akshat}. Among these, certain families of quantum machine learning (QML) models are efficiently simulatable by classical computers. Such quantum-\emph{inspired} models have garnered significant interest, as they allow quantum machine learning solutions to be developed which do not require quantum computers to be fully-fledged but which can benefit from them when they are. Foremost among these are tensor networks, which have demonstrated the ability to successfully solve machine learning tasks~\cite{stoudenmire_supervised_2016}. An alternative family are quantum-inspired parametrisations of \emph{orthogonal} neural networks~\cite{landman2022quantum, cherrat2022quantum}. Orthogonal neural networks (OrthoNNs) are a class of machine learning models that enforce orthogonality constraints on weight matrices. They have emerged as a promising tool due to their inherent stability, reduced risk of overfitting, and improved generalisation capabilities. The quantum-inspired versions of these networks inherit these positive features, but also can be implemented in a fully quantum mode, potentially delivering speedups for orthogonal neural networks at larger scales.

In this work, we demonstrate the effectiveness of the Bayesian learning paradigm to train these quantum and quantum-inspired OrthoNN parametrisations. Bayesian learning is a probabilistic framework that facilitates principled uncertainty quantification and robust decision-making. By combining the structural advantages of OrthoNNs with the probabilistic strengths of Bayesian inference, we introduce Bayesian (quantum) orthogonal neural networks (BONNs) as a framework for machine learning that ensures efficient training while providing uncertainty-aware predictions.

As an example application, we apply these BONNs to anomaly detection, a critical task in various industrial domains such as manufacturing, energy systems, and cybersecurity. Detecting anomalies involves identifying patterns or behaviours that deviate significantly from normal operations, which is essential for ensuring safety, efficiency, and reliability. Traditional anomaly detection methods often struggle with noisy data, complex patterns, and the need for high interpretability—challenges that BONNs are uniquely positioned to address.

We explore the efficacy of BONNs in anomaly detection through an industrially relevant use case, demonstrating their ability to identify anomalies with high precision and confidence. By focusing on anomalies in three-dimensional (3D) objects, we also incorporate orthogonal layers into 3D convolutional neural networks and propose the 3D orthogonal convolutional NN (OrthoConv3D). We compare the efficacy of this architecture to basic (orthogonal) feedforward networks.

\section{Background} 
\label{Sec:Background}

\begin{figure*}
    \centering
    \includegraphics[width=0.65\linewidth]{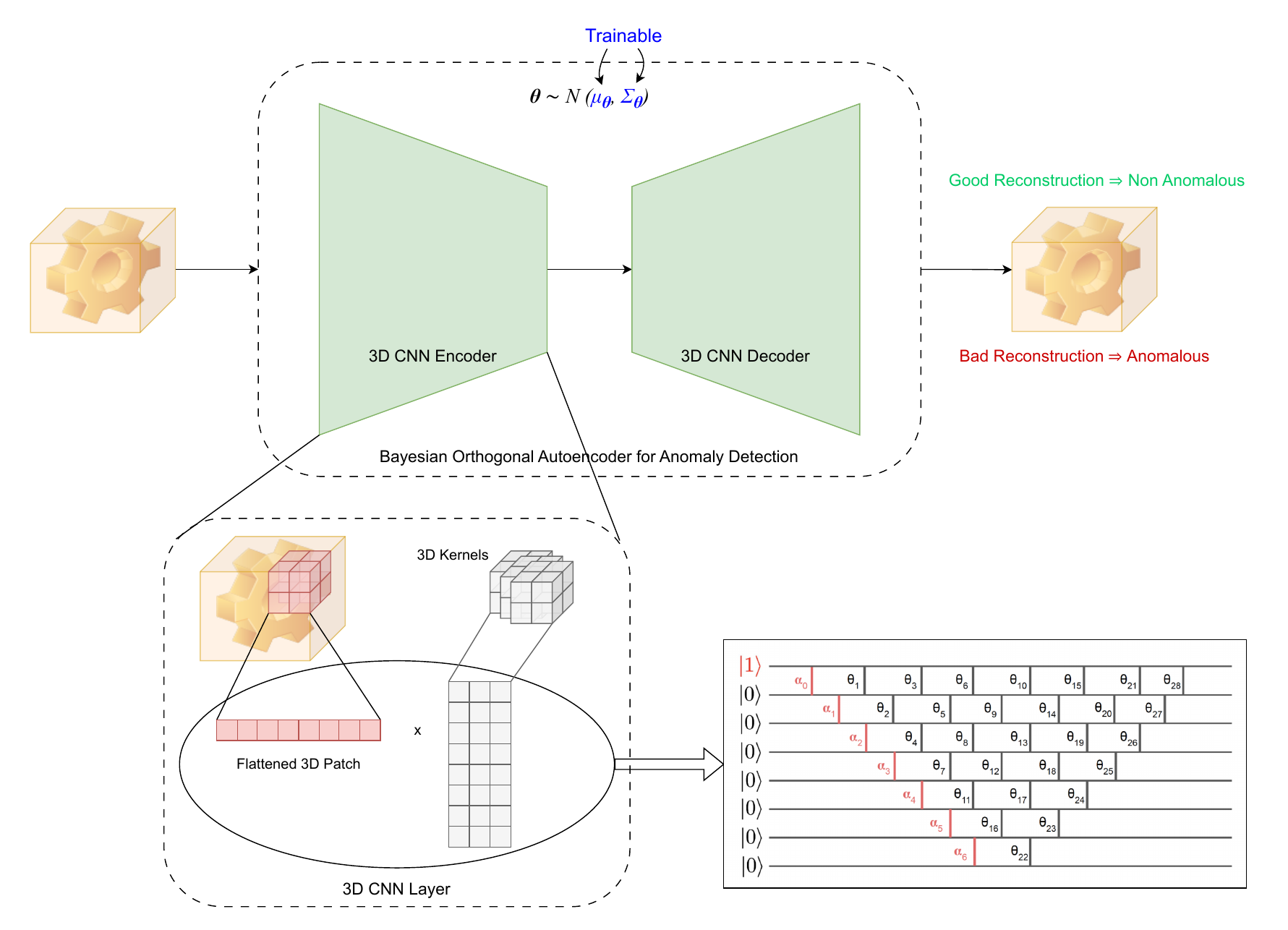}
    \caption{
    \textsf{
    \textbf{Overview of the proposed anomaly detection pipeline using orthogonal neural networks.}\\
    The 3D object (data point) is passed through 3D convolutional layers of the autoencoder, which are parametrised with orthogonal weight matrices. In the autoencoder framework, an Encoder compresses information to a 3D latent space, and a Decoder attempts to reconstruct the image from it. If the reconstruction is deemed successful, the object contains no anomaly. In the orthogonal 3D convolution, 3D patches (red) of a tensor object are flattened and treated as feature vectors. The 3D CNN kernels (grey) are flattened and form a matrix modelled by ortholinear quantum circuits (or quantum-inspired simulation) and multiplied with the flattened patches. The Loader circuit encoding this feature vector depicted is the `diagonal' loader \cite{johri2021nearest}. The orthogonal matrix multiplication corresponds to flattened 3D kernels. The orthogonal circuit shown is the `pyramid' structure \cite{landman2022quantum}. In the Bayesian framework, each orthogonal circuit parameter is sampled from a distribution, e.g. Gaussian with trainable mean and variance.
    }
    }
    \label{fig:overview}
\end{figure*}

\subsection{Additive manufacturing}
\label{Sec:UseCase}

Additive Manufacturing (AM) (also known as 3D printing) techniques have the ability to create and produce complex geometries through optimised design approaches. The potential for mass and cost savings has made AM highly competitive compared to conventional manufacturing processes, driving its rapid adoption in fields such as aerospace and automotive~\cite{Intro1, gibson_additive_2021}.

\subsection{Traditional techniques for anomaly detection in additive manufacturing}
\label{subsec:ad_traditional}

Despite extensive research and technological advancements in AM processes, identifying internal defects (anomalies) remains a persistent challenge~\cite{Intro1}. These defects can arise from various factors, including non-uniform powder distribution, variations in process parameters affecting the laser beam, scanning and building strategies, and deformations during the manufacturing process \cite{gussone_interfacial_2018, yadollahi_additive_2017}. Common defects in AM parts include inclusions, pores, and lack of fusion (LoF). Inclusions result from foreign material contamination, while pores and LoF are typically caused by process-related issues such as gas bubbles trapped during melt pool solidification and insufficient energy from the melting beam, respectively. These defects significantly impact the mechanical properties and overall quality of the 3D-printed components~\cite{greitemeier_fatigue_2017, cersullo_effect_2022, greitemeier2017fatigue, cersullo2022effect} 

A common family of approaches to identify, analyse, and classify defects are non-destructive testing (NDT) methods. These methods enable manufacturers to determine whether a part meets predefined industrial acceptance criteria based on defect size, shape, type, and distribution patterns. Among the various NDT methods, computed tomography (CT) is one of the most widely used due to its ability to provide detailed information about both the outer geometrical features and the internal structure of the part. CT scans offer a three-dimensional view, enabling comprehensive defect analysis~\cite{bordekar_explainable_2025, milcke2024exploring, bordekar2025explainable}. However, traditional CT inspection processes often involve manual analysis by human experts, which can be time-consuming, labour-intensive, and prone to inaccuracies~\cite{kiefel_computed_2018, kiefel2018computed}.

\subsection{Machine learning for anomaly detection in additive manufacturing}
\label{subsec:supervised_ad}

To address these challenges, computer vision-based methods have recently been applied to automate defect detection in CT scans~\cite{ong_solder_2008, chou_automatic_1997}, which can identify defects more efficiently and consistently. Using such automated approaches has shown promise post-printing, though their performance is often hindered by artefacts inherent to CT imaging, such as beam hardening, scatter, Poisson noise, and ringing effects and are often both time and cost intensive~\cite{boas2012ct}. Besides the cost and time constraints, the adoption of additive manufacturing (AM) for critical aviation components is also limited by the capabilities of computed tomography (CT) machines. These machines can only inspect objects within a specific size range, restricting the geometry and dimensions of AM parts that can be effectively screened. As a result, this limitation has impeded the broader integration of AM components in aviation applications~\cite{milcke2024exploring, dissertation}. 

Therefore, to broaden the application of AM  parts in critical structures, especially in aviation sectors, while simultaneously reducing production and inspection costs, implementing an online monitoring (OM) system for the manufacturing process could be highly beneficial~\cite{milcke2024exploring, dissertation}. Such a system could enable real-time quality assessment, minimising the reliance on costly post-production inspections and enhancing the reliability of such AM components. In such a method, online monitoring data can often be collected in the form of time series containing measurements of melt-pool radiation intensity measured at each point, during the printing phase of the parts. This time-series data is then mapped to a 3D representation of the printed parts, which can then be further processed and used for automatic defect detection using current AI methods. Such in-process assessment may reduce reliance on costly and time-consuming post-production inspections, facilitating the production of larger and more complex components that exceed the size limitations of computed tomography (CT) scanning~\cite{milcke2024exploring, dissertation}.

Online monitoring of additive manufacturing (AM) generates vast amounts of data that must also be analysed for potential deviations and defect detection. AI methods have been shown to outperform traditional image processing techniques in such tasks~\cite{milcke2024exploring, dissertation, masci2012steel, tang2021anomaly}. Most commonly, supervised learning techniques are applied~\cite{masci2012steel, tang2021anomaly}. In many of these approaches, a semantic segmentation is performed on the 3D voxels of the printed part data. The brightness or intensity of the voxels is analysed with an appropriate threshold selected to segment the pixels/voxels as defective or not~\cite{gong2019micro}. 
% This technique also facilitates the precise localization of various defects. For example, \cite{fuchs2019generating} uses the U-NET architecture to detect defects (for example, pores or shrinkage) in aluminium casting parts. Ref.~\cite{gobert2020porosity} combined convolutional neural networks (CNNs) with thresholding techniques for automatic porosity segmentation in X-ray images or detect defects in online monitoring data, further improving quality control in AM~\cite{milcke2024exploring, dissertation}.

\subsection{Autoencoders for anomaly detection in additive manufacturing}
\label{subsec:ae_for_ad}

However, they outperform other traditional methods, supervised approaches have significant data requirements~\cite{singh2020explainable} and may require expert involvement for labelling high-quality training datasets, which is often expensive but ultimately determines model performance~\cite{fuchs2019generating, dissertation}. Additionally, supervised methods often rely heavily on well-balanced labelled datasets. In AM applications, however, defects occur infrequently, leading to an imbalance between normal and abnormal data, which poses a significant challenge~\cite{tan2023deep, lee2025autonomous} to establish correlations between online monitoring data and the actual presence or absence of defects.

To address this limitation, various unsupervised learning approaches have been explored to reduce the need for labelled data. These include auto-encoders (AE)~\cite{zhou2017anomaly}, variational auto-encoders (VAE)~\cite{an2015variational}, and generative adversarial networks (GANs) \cite{hu2020unsupervised, xia2022gan}. Among these, VAEs are the most commonly used for similar applications. Autoencoders can offer guaranteed convergence and simpler training procedures compared to GANs~\cite{lee2025autonomous}, and variational versions can learn the underlying data distribution.

However, current unsupervised approaches are still far from perfect. For example, even though applying such methods on online monitoring data mitigates time and cost issues from the use of CT-scan data, such data also contains certain artefacts, noise, and other distortions. Plus, the reliability of model predictions also extends beyond their accuracy. Especially in the context of non-destructive testing (NDT), where the reliability of such model prediction is crucial for building trust in both the AI model and the overall methodology~\cite{bordekar2025explainable}. To address this, some recent work enhances model robustness by estimating the \emph{uncertainty} of predictions using Bayesian methods or other techniques. These techniques improve interpretability and the models' ability to handle noise and artefacts, resulting in more reliable defect detection and stronger predictive performance~\cite{bordekar2025explainable, milcke2024exploring}. It is for these reasons that we propose to use the combination of variational autoencoders and Bayesian learning for our quantum(-inspired) anomaly detection pipeline.

\subsection{Orthogonal Neural Networks} 
\label{SubSec:OrthoNNs}

Fully connected neural networks layers, produce an output, \(\boldsymbol{y} := \sigma(W\boldsymbol{x} + \boldsymbol{b})\) given an input \(\boldsymbol{x}\in \mathbb{R}^{d}\). We have \(\mathbf{W} \in \mathbb{R}^{d \times d}\), a bias vector, \(\boldsymbol{b}\in \mathbb{R}^{d}\), and a non-linear activation function, \(\sigma(\cdot)\). A particular family of neural network called \emph{orthogonal} neural networks (OrthoNNs) add an orthogonality constraint on \(W\) - \(\mathbf{W}^{\top}\mathbf{W} = \mathds{1}_{d}\). This feature can provide benefits in many aspects of learning, for example, in training by avoiding exploding or vanishing gradients~\cite{jia2017improving, xiao2017fashion, saxe2013exact} during backpropagation. Orthogonality can also grant theoretical and practical benefits in generalisation error~\cite{jia2019orthogonal}. One can also construct composite architectures beyond simple linear layers, such as orthogonal convolutions~\cite{wang2020orthogonal}.

While orthogonality constraints add stability and learning benefits, enforcing them in practice can be difficult. The forward pass of an OrthoNN has the same complexity as a non-orthogonal fully connected layer, but backpropagation is more costly. This is due to the orthogonality-breaking nature of gradient updates which requires that the weight matrix must be constantly re-orthogonalised. Exact orthogonalisation is expensive - for example exponentiation of the optimised matrix logarithm requires \(\mathcal{O}(d^3)\) complexity, and also may be numerically unstable with large-scale matrices. On the other hand, techniques such as singular value bounding (SVB)~\cite{jia2019orthogonal} are more efficient but only enforce \emph{approximate} orthogonality, losing some theoretical benefits.

\subsubsection{Orthogonal Quantum Neural Networks} \label{sec:ortho_QNNs}

To address this, Ref.~\cite{landman2022quantum} introduced a hybrid classical-quantum approach for implementing orthogonal neural networks. Here, the orthogonal layer is represented as a quantum circuit with trainable parameters and decomposed using Givens rotations. Such OrthoNNs may be implemented in a \emph{quantum-inspired} mode, and hence scalable to large problem sizes independent of quantum hardware. Also, they may deliver a small polynomial speedup in implementing OrthoNNs directly on quantum computers, as they develop. 

Practically speaking, these circuits contain Hamming-weight preserving operations, known as reconfigurable beam splitter (RBS) gates as in Eq.~\ref{eqn:rbs_gate_defn}, acting between two qubits \(i, j\).
\begin{multline} \label{eqn:rbs_gate_defn}
    \mathsf{RBS}_{i, j}(\theta) := e^{-i\frac{\theta}{2}\left(Y_i \otimes X_j - X_i \otimes Y_j\right)} \\
    = \left(
    \begin{array}{cccc}
    1 & 0 & 0 & 0 \\
    0 & \cos(\theta) & -\sin(\theta) & 0 \\
    0 & \sin(\theta) & \cos(\theta) & 0 \\
    0 & 0 & 0 & 1 
    \end{array}
    \right)
\end{multline}

These are a subset of the larger family of \emph{subspace-preserving} quantum circuits. For these circuits, input data vectors (normalised) are encoded in the Hamming-weight \(1\) (unary) subspace of a quantum state, \(\ket{
\boldsymbol{x}} = \sum_{j}x_j \ket{\boldsymbol{e}_j}\), where $\textsf{HW}(\boldsymbol{e}_j) = 1$ is a Hamming-weight \(1\) computational basis state. Due to the Hamming-weight preserving nature of the above gates~\eqref{eqn:rbs_gate_defn}, the output state \(\ket{\boldsymbol{y}} = \sum_{j}y_j \ket{\boldsymbol{e}_j}\) will also be supported exclusively on the Hamming-weight \(1\) subspace. The vector \(\boldsymbol{y}\) can be extracted efficiently via \(\ell_{\infty}\) tomography~\cite{landman2022quantum}, and related to the original vector exactly by an orthogonal matrix, \(\boldsymbol{y} = O\boldsymbol{x}\). The components of \(O\) are related to the angles, \(\boldsymbol{\theta}\) in the circuit unitary, \(U_O(\boldsymbol{\theta})\). The exact dependency will depend on the structure of the \(\textsf{RBS}\) gates within \(U_O(\boldsymbol{\theta})\). 

The above circuits are efficiently classically simulatable due to the restriction to the unary subspace. To increase the classical simulation difficulty of these circuits, we can generalise to higher Hamming-weight data encodings (computational basis states with Hamming weight \(\mathsf{HW}(\boldsymbol{e}_k) = k > 1\)). This is due to the increase in subspace dimension needing to be simulated as \(\binom{n}{k}\). For example, \emph{compound} layers or compound neural networks~\cite{thakkar2024improved} can operate on all \(k < \frac{n}{2}\) Hamming-weights simultaneously using circuits containing \emph{Fermionic} beam splitter gates (\textsf{FBS}) - generalisations of Eq.~\ref{eqn:rbs_gate_defn} which account for the parity between qubits \(i, j\)~\cite{kerenidis2022quantum}.  The effective action on these higher HW subspaces is of compound matrices of increasing order for \textsf{FBS} circuits. Using the above orthogonal primitives, we can construct various advanced models like quantum vision transformers~\cite{cherrat2022quantum}, and quantum Fourier networks~\cite{jain2024quantum}. In this work, we will add three-dimensional convolutions to this list. Importantly, these models can be free from barren plateaus~\cite{monbroussou2023trainability, raj2023quantum, fontana2023adjoint, cerezo2023does}, which is an essential feature in training non-classically simulatable versions on quantum hardware.

\begin{figure}[ht]
\centering
\begin{subfigure}{0.6\columnwidth}
  \centering
  \includegraphics[width=\linewidth]{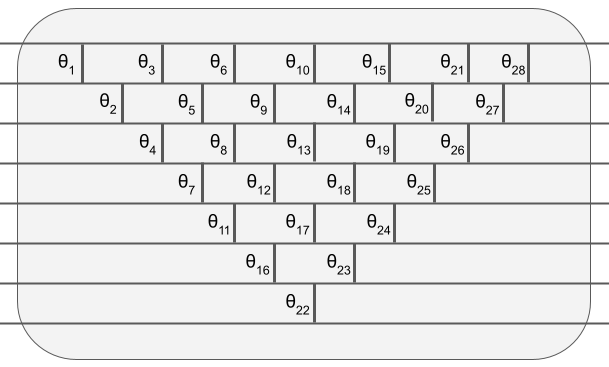}
  \caption{}
  \label{fig:QONNcircuit}
\end{subfigure}%
\begin{subfigure}{0.35\columnwidth}
  \centering
  \includegraphics[width=\linewidth]{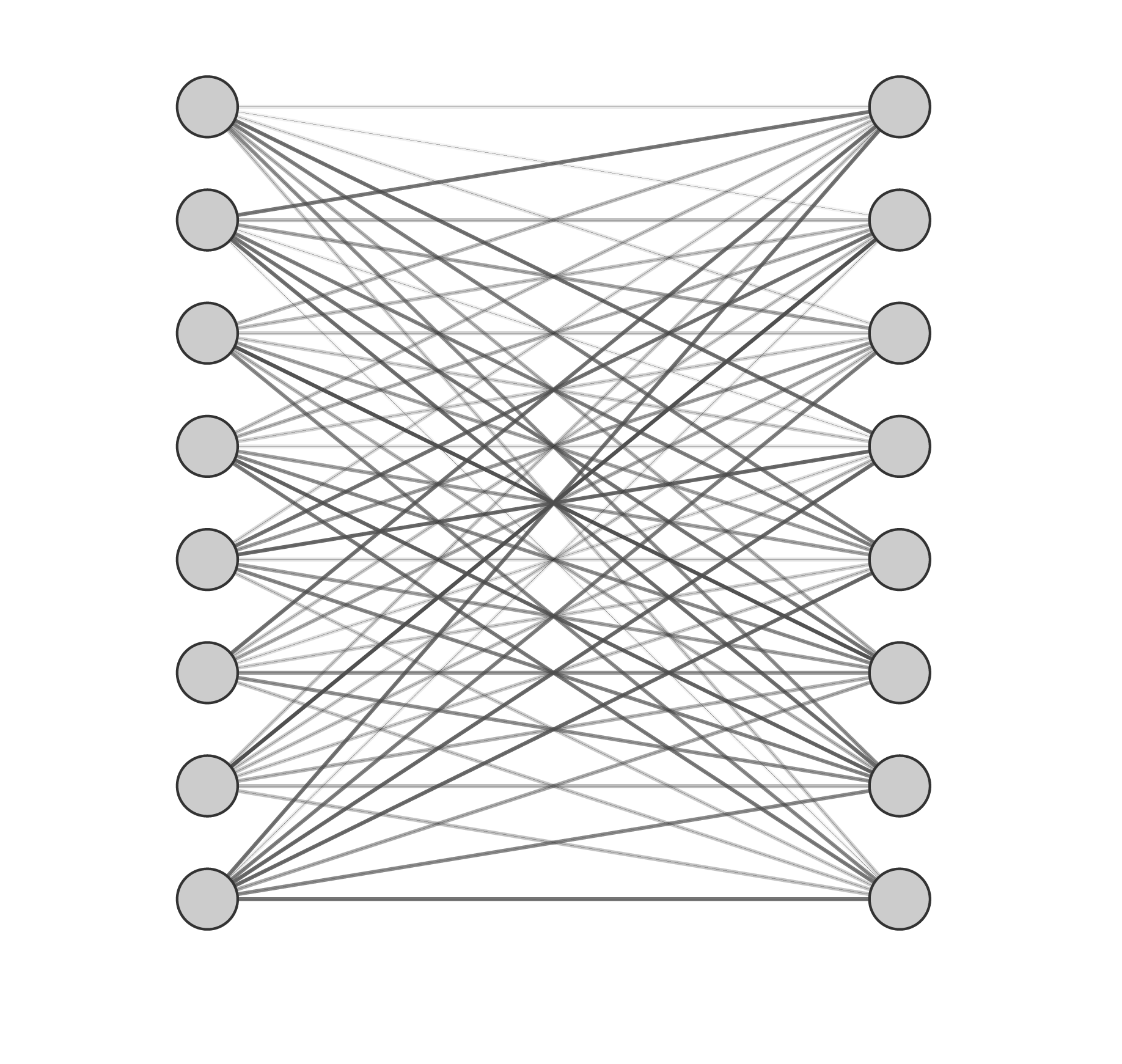}
  \caption{}
  \label{fig:8-8-nn}
\end{subfigure}
\caption{ (a) Parametrised quantum circuit for an $8 \times 8$ quantum orthogonal layer. Each vertical line corresponds to an $RBS$ gate with its angle parameter. And (b), a classical (non-orthogonal) neural network $8 \times 8$ layer. Courtesy~\cite{landman2022quantum}.}
\label{fig:QONNcircuit_comparison}
\end{figure}

\begin{table*}[ht!]
\centering
\resizebox{0.8\textwidth}{!}{%
\begin{tabular}{||l|c|c|c|c|c|c||}
\hline
Models     & Precision      & Estimated Calibration Error & Recall         & F1-Score       & SDA & LuDA \\ \hline
FNN (PE)   & \textbf{74.09}          & 0.257                       & 71.47          & 72.29          & 4   & 22   \\
FNN (Bayesian)   & 71.68          & \textbf{0.221}              & 71.03          & 71.35          & 4   & 15   \\
QFNN (PE)     & 73.95  & 0.251                    & \textbf{71.78}  & \textbf{72.84}  & 4  & 17 \\
QFNN (Bayesian)    & 72.89  & \textbf{0.217} & 71.65  & 72.26  & 4  & 17 \\
\hline
\end{tabular}%
}
\caption{
\textsf{\small
\textbf{Anomaly detection with Bayesian learning - feedforward}. \\
Comparing Bayesian and non-Bayesian (i.e. standard gradient descent on a classification loss function) learning for classical (FNN) and quantum (QFNN) feedforward networks in an autoencoder pipeline. We also show the smallest detected anomaly (SDA) and the largest \emph{un}detected anomaly (LuDA) for each model. Bayesian methods are clearly superior when Estimated Calibration Error is the metric of interest. Hybrid quantum (orthogonal) networks also generally outperform their vanilla counterparts, highlighting the benefits of orthogonality for anomaly detection.
}
}
\label{tab:fnn_results}
\end{table*}

\begin{figure*}[ht!]
  \centering
  %--- Top Row: ECE, SDA, LuDA ---
  % ECE subplot:
  \begin{subfigure}[b]{0.32\textwidth}
    \centering
    \begin{tikzpicture}
      \begin{axis}[
          ybar,
          xtick={0,1,2,3,4,5,6,7},
          xticklabels={}, 
          ylabel={\sffamily ECE},
          ymin=2.1, ymax=2.7,
          grid=both,
          minor grid style={dashed,gray!20},
          major grid style={solid,gray!50},
          width=5.5cm,
          height=3cm,
          bar width=12pt,
          every axis plot/.append style={bar shift=0pt},
          cycle list={{}}
      ]
        % Bars:
        \addplot[fill=TangoBlue, draw=black, fill opacity=0.4]    coordinates {(0, 2.57)};
        \addplot[fill=TangoCyan, draw=black, fill opacity=0.4]  coordinates {(1, 2.51)};
        \addplot[fill=TangoRed, draw=black, fill opacity=0.4]   coordinates {(2, 2.21)};
        \addplot[fill=TangoOrange, draw=black, fill opacity=0.4]  coordinates {(3, 2.17)};
        \addplot[fill=TangoPurple, draw=black, fill opacity=0.4]     coordinates {(4, 2.32)};
        \addplot[fill=TangoMagenta, draw=black, fill opacity=0.4]  coordinates {(5, 2.27)};
        \addplot[fill=TangoGreen, draw=black, fill opacity=0.4] coordinates {(6, 2.41)};
        \addplot[fill=TangoButter, draw=black, fill opacity=0.4]    coordinates {(7, 2.41)};
        % Markers:
        \addplot[only marks, mark=square*, mark size=3.5, thick, color=TangoBlue, forget plot]     coordinates {(0, 2.57)};
        \addplot[only marks, mark=triangle*, mark size=3.5, thick, color=TangoCyan, forget plot]   coordinates {(1, 2.51)};
        \addplot[only marks, mark=+, mark options={line width=2, mark size=5}, thick, color=TangoRed, forget plot]           coordinates {(2, 2.21)};
        \addplot[only marks, mark=diamond*, mark size=3.5, thick, color=TangoOrange, forget plot]      coordinates {(3, 2.17)};
        \addplot[only marks, mark=pentagon*, mark size=3.5, thick, color=TangoPurple, forget plot]        coordinates {(4, 2.32)};
        \addplot[only marks, mark=star, mark size=3.5, thick, color=TangoMagenta, forget plot]          coordinates {(5, 2.27)};
        \addplot[only marks, mark=oplus*, mark size=3.5, thick, color=TangoGreen, forget plot]       coordinates {(6, 2.41)};
        \addplot[only marks, mark=otimes*, mark size=3.5, thick, color=TangoButter, forget plot]          coordinates {(7, 2.41)};
      \end{axis}
    \end{tikzpicture}
      \vspace{-0.4cm}
    \caption{ECE (\(\times 10^{-1}\)) (\(\boldsymbol{\downarrow}\))}
  \end{subfigure}
  % \hfill
  % SDA subplot:
  \begin{subfigure}[b]{0.32\textwidth}
    \centering
    \begin{tikzpicture}
      \begin{axis}[
          ybar,
          xtick={0,1,2,3,4,5,6,7},
          xticklabels={},
          ylabel={\sffamily SDA},
          ymin=2, ymax=5,
          grid=both,
          minor grid style={dashed,gray!20},
          major grid style={solid,gray!50},
          width=5.5cm,
          height=3cm,
          bar width=12pt,
          every axis plot/.append style={bar shift=0pt},
          cycle list={{}}
      ]
        % All SDA values are 4.
        \addplot[fill=TangoBlue, draw=black, fill opacity=0.4]    coordinates {(0,4)};
        \addplot[fill=TangoCyan, draw=black, fill opacity=0.4]  coordinates {(1,4)};
        \addplot[fill=TangoRed, draw=black, fill opacity=0.4]   coordinates {(2,4)};
        \addplot[fill=TangoOrange, draw=black, fill opacity=0.4]  coordinates {(3,4)};
        \addplot[fill=TangoPurple, draw=black, fill opacity=0.4]     coordinates {(4,4)};
        \addplot[fill=TangoMagenta, draw=black, fill opacity=0.4]  coordinates {(5,4)};
        \addplot[fill=TangoGreen, draw=black, fill opacity=0.4] coordinates {(6,4)};
        \addplot[fill=TangoButter, draw=black, fill opacity=0.4]    coordinates {(7,4)};
        % Markers:
        \addplot[only marks, mark=square*, mark size=3.5, thick, color=TangoBlue, forget plot]     coordinates {(0,4)};
        \addplot[only marks, mark=triangle*, mark size=3.5, thick, color=TangoCyan, forget plot]   coordinates {(1,4)};
        \addplot[only marks, mark=+, mark options={line width=2, mark size=5}, thick, color=TangoRed, forget plot]           coordinates {(2,4)};
        \addplot[only marks, mark=diamond*, mark size=3.5, thick, color=TangoOrange, forget plot]      coordinates {(3,4)};
        \addplot[only marks, mark=pentagon*, mark size=3.5, thick, color=TangoPurple, forget plot]        coordinates {(4,4)};
        \addplot[only marks, mark=star, mark size=3.5, thick, color=TangoMagenta, forget plot]          coordinates {(5,4)};
        \addplot[only marks, mark=oplus*, mark size=3.5, thick, color=TangoGreen, forget plot]       coordinates {(6,4)};
        \addplot[only marks, mark=otimes*, mark size=3.5, thick, color=TangoButter, forget plot]          coordinates {(7,4)};
      \end{axis}
    \end{tikzpicture}
      \vspace{-0.4cm}
    \caption{SDA}
  \end{subfigure}
  % \hfill
  % LuDA subplot:
  \begin{subfigure}[b]{0.32\textwidth}
    \centering
    \begin{tikzpicture}
      \begin{axis}[
          ybar,
          xtick={0,1,2,3,4,5,6,7},
          xticklabels={},
          ylabel={\sffamily LuDA},
          ymin=14, ymax=24,
          grid=both,
          minor grid style={dashed,gray!20},
          major grid style={solid,gray!50},
          width=5.5cm,
          height=3cm,
          bar width=12pt,
          every axis plot/.append style={bar shift=0pt},
          cycle list={{}}
      ]
        \addplot[fill=TangoBlue, draw=black, fill opacity=0.4]    coordinates {(0,22)};
        \addplot[fill=TangoCyan, draw=black, fill opacity=0.4]  coordinates {(1,17)};
        \addplot[fill=TangoRed, draw=black, fill opacity=0.4]   coordinates {(2,15)};
        \addplot[fill=TangoOrange, draw=black, fill opacity=0.4]  coordinates {(3,17)};
        \addplot[fill=TangoPurple, draw=black, fill opacity=0.4]     coordinates {(4,15)};
        \addplot[fill=TangoMagenta, draw=black, fill opacity=0.4]  coordinates {(5,15)};
        \addplot[fill=TangoGreen, draw=black, fill opacity=0.4] coordinates {(6,15)};
        \addplot[fill=TangoButter, draw=black, fill opacity=0.4]    coordinates {(7,17)};
        % Markers:
        \addplot[only marks, mark=square*, mark size=3.5, thick, color=TangoBlue, forget plot]     coordinates {(0,22)};
        \addplot[only marks, mark=triangle*, mark size=3.5, thick, color=TangoCyan, forget plot]   coordinates {(1,17)};
        \addplot[only marks, mark=+, mark options={line width=2, mark size=5}, thick, color=TangoRed, forget plot]           coordinates {(2,15)};
        \addplot[only marks, mark=diamond*, mark size=3.5, thick, color=TangoOrange, forget plot]      coordinates {(3,17)};
        \addplot[only marks, mark=pentagon*, mark size=3.5, thick, color=TangoPurple, forget plot]        coordinates {(4,15)};
        \addplot[only marks, mark=star, mark size=3.5, thick, color=TangoMagenta, forget plot]          coordinates {(5,15)};
        \addplot[only marks, mark=oplus*, mark size=3.5, thick, color=TangoGreen, forget plot]       coordinates {(6,15)};
        \addplot[only marks, mark=otimes*, mark size=3.5, thick, color=TangoButter, forget plot]          coordinates {(7,17)};
      \end{axis}
    \end{tikzpicture}
      \vspace{-0.4cm}
    \caption{LuDA}
  \end{subfigure}\\[0.5cm]
  %--- Bottom Row: Precision, Recall, F1-Score ---
  % Precision subplot:
  \begin{subfigure}[b]{0.32\textwidth}
    \centering
    \begin{tikzpicture}
      \begin{axis}[
          ybar,
          xtick={0,1,2,3,4,5,6,7},
          xticklabels={},
          ylabel={\sffamily Precision},
          ymin=71, ymax=76,
          grid=both,
          minor grid style={dashed,gray!20},
          major grid style={solid,gray!50},
          width=5.5cm,
          height=3cm,
          bar width=12pt,
          every axis plot/.append style={bar shift=0pt},
          cycle list={{}}
      ]
         \addplot[fill=TangoBlue, draw=black, fill opacity=0.4]    coordinates {(0,73.14)};
         \addplot[fill=TangoCyan, draw=black, fill opacity=0.4]  coordinates {(1,73.95)};
         \addplot[fill=TangoRed, draw=black, fill opacity=0.4]   coordinates {(2,71.68)};
         \addplot[fill=TangoOrange, draw=black, fill opacity=0.4]  coordinates {(3,72.89)};
         \addplot[fill=TangoPurple, draw=black, fill opacity=0.4]     coordinates {(4,74.46)};
         \addplot[fill=TangoMagenta, draw=black, fill opacity=0.4]  coordinates {(5,74.25)};
         \addplot[fill=TangoGreen, draw=black, fill opacity=0.4] coordinates {(6,74.97)};
         \addplot[fill=TangoButter, draw=black, fill opacity=0.4]    coordinates {(7,74.33)};
         % Markers:
         \addplot[only marks, mark=square*, mark size=3.5, thick, color=TangoBlue, forget plot]    coordinates {(0,73.14)};
         \addplot[only marks, mark=triangle*, mark size=3.5, thick, color=TangoCyan, forget plot]  coordinates {(1,73.95)};
         \addplot[only marks,mark=+, mark options={line width=2, mark size=5}, thick, color=TangoRed, forget plot]          coordinates {(2,71.68)};
         \addplot[only marks, mark=diamond*, mark size=3.5, thick, color=TangoOrange, forget plot]     coordinates {(3,72.89)};
         \addplot[only marks, mark=pentagon*, mark size=3.5, thick, color=TangoPurple, forget plot]       coordinates {(4,74.46)};
         \addplot[only marks, mark=star, mark size=3.5, thick, color=TangoMagenta, forget plot]         coordinates {(5,74.25)};
         \addplot[only marks, mark=oplus*, mark size=3.5, thick, color=TangoGreen, forget plot]      coordinates {(6,74.97)};
         \addplot[only marks, mark=otimes*, mark size=3.5, thick, color=TangoButter, forget plot]         coordinates {(7,74.33)};
      \end{axis}
    \end{tikzpicture}
    \vspace{-0.4cm}
    \caption{Precision (\(\boldsymbol{\uparrow}\))}
  \end{subfigure}
  % \hfill
  % Recall subplot:
  \begin{subfigure}[b]{0.32\textwidth}
    \centering
    \begin{tikzpicture}
      \begin{axis}[
          ybar,
          xtick={0,1,2,3,4,5,6,7},
          xticklabels={},
          ylabel={\sffamily Recall},
          ymin=70, ymax=73,
          grid=both,
          minor grid style={dashed,gray!20},
          major grid style={solid,gray!50},
          width=5.5cm,
          height=3cm,
          bar width=12pt,
          every axis plot/.append style={bar shift=0pt},
          cycle list={{}}
      ]
         \addplot[fill=TangoBlue, draw=black, fill opacity=0.4]    coordinates {(0,71.47)};
         \addplot[fill=TangoCyan, draw=black, fill opacity=0.4]  coordinates {(1,71.78)};
         \addplot[fill=TangoRed, draw=black, fill opacity=0.4]   coordinates {(2,71.03)};
         \addplot[fill=TangoOrange, draw=black, fill opacity=0.4]  coordinates {(3,71.65)};
         \addplot[fill=TangoPurple, draw=black, fill opacity=0.4]     coordinates {(4,71.89)};
         \addplot[fill=TangoMagenta, draw=black, fill opacity=0.4]  coordinates {(5,72.07)};
         \addplot[fill=TangoGreen, draw=black, fill opacity=0.4] coordinates {(6,72.13)};
         \addplot[fill=TangoButter, draw=black, fill opacity=0.4]    coordinates {(7,72.21)};
         % Markers:
         \addplot[only marks, mark=square*, mark size=3.5, thick, color=TangoBlue, forget plot]    coordinates {(0,71.47)};
         \addplot[only marks, mark=triangle*, mark size=3.5, thick, color=TangoCyan, forget plot]  coordinates {(1,71.78)};
         \addplot[only marks, mark=+, mark options={line width=2, mark size=5}, thick, color=TangoRed, forget plot]          coordinates {(2,71.03)};
         \addplot[only marks, mark=diamond*, mark size=3.5, thick, color=TangoOrange, forget plot]     coordinates {(3,71.65)};
         \addplot[only marks, mark=pentagon*, mark size=3.5, thick, color=TangoPurple, forget plot]       coordinates {(4,71.89)};
         \addplot[only marks, mark=star, mark size=3.5, thick, color=TangoMagenta, forget plot]         coordinates {(5,72.07)};
         \addplot[only marks, mark=oplus*, mark size=3.5, thick, color=TangoGreen, forget plot]      coordinates {(6,72.13)};
         \addplot[only marks, mark=otimes*, mark size=3.5, thick, color=TangoButter, forget plot]         coordinates {(7,72.21)};
      \end{axis}
    \end{tikzpicture}
      \vspace{-0.4cm}
    \caption{Recall (\(\boldsymbol{\uparrow}\))}
  \end{subfigure}
  % \hfill
  % F1-Score subplot:
  \begin{subfigure}[b]{0.32\textwidth}
    \centering
    \begin{tikzpicture}
      \begin{axis}[
          ybar,
          xtick={0,1,2,3,4,5,6,7},
          xticklabels={},
          ylabel={\sffamily F1-Score},
          ymin=71, ymax=74,
          grid=both,
          minor grid style={dashed,gray!20},
          major grid style={solid,gray!50},
          width=5.5cm,
          height=3cm,
          bar width=12pt,
          every axis plot/.append style={bar shift=0pt},
          cycle list={{}}
      ]
         \addplot[fill=TangoBlue, draw=black, fill opacity=0.4]    coordinates {(0,72.29)};
         \addplot[fill=TangoCyan, draw=black, fill opacity=0.4]  coordinates {(1,72.84)};
         \addplot[fill=TangoRed, draw=black, fill opacity=0.4]   coordinates {(2,71.35)};
         \addplot[fill=TangoOrange, draw=black, fill opacity=0.4]  coordinates {(3,72.26)};
         \addplot[fill=TangoPurple, draw=black, fill opacity=0.4]     coordinates {(4,73.15)};
         \addplot[fill=TangoMagenta, draw=black, fill opacity=0.4]  coordinates {(5,73.14)};
         \addplot[fill=TangoGreen, draw=black, fill opacity=0.4] coordinates {(6,73.52)};
         \addplot[fill=TangoButter, draw=black, fill opacity=0.4]    coordinates {(7,73.25)};
         % Markers:
         \addplot[only marks, mark=square*, mark size=3.5, thick, color=TangoBlue, forget plot]    coordinates {(0,72.29)};
         \addplot[only marks, mark=triangle*, mark size=3.5, thick, color=TangoCyan, forget plot]  coordinates {(1,72.84)};
         \addplot[only marks, mark=+, mark options={line width=2, mark size=5}, thick, color=TangoRed, forget plot]           coordinates {(2,71.35)};
         \addplot[only marks, mark=diamond*, mark size=3.5, thick, color=TangoOrange, forget plot]     coordinates {(3,72.26)};
         \addplot[only marks, mark=pentagon*, mark size=3.5, thick, color=TangoPurple, forget plot]       coordinates {(4,73.15)};
         \addplot[only marks, mark=star, mark size=3.5, thick, color=TangoMagenta, forget plot]         coordinates {(5,73.14)};
         \addplot[only marks, mark=oplus*, mark size=3.5, thick, color=TangoGreen, forget plot]      coordinates {(6,73.52)};
         \addplot[only marks, mark=otimes*, mark size=3.5, thick, color=TangoButter, forget plot]         coordinates {(7,73.25)};
      \end{axis}
    \end{tikzpicture}
      \vspace{-0.4cm}
    \caption{F1-Score (\(\boldsymbol{\uparrow}\))}
  \end{subfigure}
  \begin{subfigure}[b]{\textwidth}
  \centering
  \captionsetup{labelformat=empty}
  \begin{tikzpicture}[baseline=(current bounding box.center)]
    \matrix (m) [matrix of nodes, 
                 nodes={anchor=west, align=center},
                 column sep=0.05cm, row sep=0.05cm]{
      % First row: PE and Bayesian entries
      {\tikz \draw[mark=square*, mark size=5, color=TangoBlue] plot coordinates {(0,0)};} & {\sffamily\footnotesize FNN (PE)} & &&
      {\tikz \draw[mark=triangle*, mark size=5, color=TangoCyan] plot coordinates {(0,0)};} & {\sffamily\footnotesize QFNN (PE)} & &&
      {\tikz \draw[mark=+, mark options={line width=2, mark size=5}, color=TangoRed] plot coordinates {(0,0)};} & {\sffamily\footnotesize FNN (Bayesian)} & &&
      {\tikz \draw[mark=diamond*, mark size=5, color=TangoOrange] plot coordinates {(0,0)};} & {\sffamily\footnotesize QFNN (Bayesian)} \\
      % Second row: MCD and Ensemble entries
      {\tikz \draw[mark=pentagon*, mark size=5, color=TangoMagenta] plot coordinates {(0,0)};} & {\sffamily\footnotesize FNN (MCD)} & &&
      {\tikz \draw[mark=star, mark options={line width=2, mark size=5}, color=TangoPurple] plot coordinates {(0,0)};} & {\sffamily\footnotesize QFNN  (MCD)} & &&
      {\tikz \draw[mark=oplus*, mark size=5, color=TangoGreen] plot coordinates {(0,0)};} & {\sffamily\footnotesize FNN (Ensemble)} & &&
      {\tikz \draw[mark=otimes*, mark size=5, color=TangoButter] plot coordinates {(0,0)};} & {\sffamily\footnotesize QFNN (Ensemble)} \\
    };
  \end{tikzpicture}
\end{subfigure}
  \caption{\small \textsf{
  \textbf{Comparing uncertainty prediction approaches with feedforward architectures}.
  Each plot shows a different metric, for each of the approaches. a) Expected Calibration Error (ECE), which is our quantifier of a robust predictor. b) The smallest detected anomaly (SDA) and largest \emph{un}detected anomaly (LuDA) for each model. Finally, standard supervised metrics, d) Precision, e) Recall and f) F1-Score. For both FNN and QFNN architectures, we test point-estimate gradient descent (PE), Bayesian learning, Monte Carlo Dropout (MCD) and Ensembling. The latter are alternative uncertainty prediction methods. Focusing on the ECE as the primary metric (lower is better), we see the hybrid quantum FNN with Bayesian training outperforms all other models.
  }
  }
  \label{fig:anomaly_results}
\end{figure*}

\subsection{Bayesian learning of neural networks}
\label{SubSec:BayesianLearning}
The typical training procedure for neural network parametrised function, \(f(\boldsymbol{\theta})\), is to find an optimal point estimate for the parameters (e.g. weights and biases), \(\boldsymbol{\theta}:= \{\mathbf{W}, \boldsymbol{b}\}\), which optimises a loss function. This formulation often struggles in estimating uncertainty/confidence regarding the predicted value. In critical domains such as medical prediction~\cite{kompa_second_2021}, the ability of the model to express uncertainty and pass to a human expert when unsure is essential for patient safety.

In Bayesian learning, instead of a point estimate, the parameters, \(\boldsymbol{\theta}\) are treated as random variables, following a (posterior) \emph{distribution}, \(\pi(\boldsymbol{\theta})\). Then, finding an optimal set of parameters can be written as a maximisation problem of these parameters, given data (for machine learning purposes) \(\pi(\boldsymbol{\theta}|\mathcal{D})\). In practice, the data is accessed via finite training/test datasets, and given by \(\mathcal{D} := \{\boldsymbol{x}_i\}_{i=1}^D, D := |\mathcal{D}|\). 

If we have access to the posterior distribution $\pi(\boldsymbol{\theta}|\mathcal{D})$, we can then estimate the output $y$ for any given input $\boldsymbol{x}$ using marginalisation as follows:
\begin{equation} \label{eqn:marginal_label_prediction_bayesian}
    p(y|\boldsymbol{x}, \mathcal{D}) = \int p(y|\boldsymbol{\theta}, \boldsymbol{x}) \pi(\boldsymbol{\theta}|\mathcal{D}) d\boldsymbol{\theta}
\end{equation}
Where we also must have some representation of the \emph{likelihood}, \(p(y|\boldsymbol{\theta}, \boldsymbol{x})\). Typically, the likelihood will be defined via the choice of model, and we can approximate \(p(y|\boldsymbol{x}, \mathcal{D})\) by sampling parameters from the posterior, and passing these through the model for a given \(\{\boldsymbol{x}, y\}\).\\

\begin{table*}[ht!]
\centering
\resizebox{0.8\textwidth}{!}{%
\begin{tabular}{||l|c|c|c|c|c|c||}
\hline
Models             & Precision      & Estimated Calibration Error & Recall         & F1-Score       & SDA & LuDA \\ \hline
3D-CNN (PE)           & \textbf{76.35}          & 0.239                       & \textbf{74.97}          & \textbf{75.65}          & 4   & 14   \\
3D-CNN (Bayesian)   & 73.98          & \textbf{0.209}              & 72.86          & 73.41          & 4   & 14   \\
3D-QCNN (PE)         & 75.49          & 0.242          & 74.12          & 74.79          & 4 & 15          \\
3D-QCNN (Bayesian)    & 73.13          & \textbf{0.224} & 72.37          & 72.74          & 4 & 17          \\ \hline
\end{tabular}%
}
\caption{
\small
\textsf{
\textbf{Anomaly detection with Bayesian learning - 3D convolution}. 
Comparing Bayesian and non-Bayesian (i.e. standard gradient descent on a classification loss function) learning for classical (3D-CNN) and quantum (3D-QCNN) 3D convolutional neural networks in an autoencoder pipeline. Again, we have the smallest detected anomaly (SDA) and the largest \emph{un}detected anomaly (LuDA) for each model. Again, Bayesian methods outperform point-estimate methods when Estimated Calibration Error (ECE) is the metric of interest. However, in this case, orthogonality does not help.
}
}
\label{tab:cnn_results_bayesian}
\end{table*}

% \sout{\textbf{Training Bayesian Neural Networks.} }
\noindent \textbf{Computing the posterior} The posterior distribution $\pi(\boldsymbol{\theta}|\mathcal{D})$ can be calculated as follows using Bayes rule:

\begin{equation} \label{eqn:bayes_rule}
\pi(\boldsymbol{\theta}|\mathcal{D}) = \frac{p(\mathcal{D}|\boldsymbol{\theta})p(\boldsymbol{\theta})}{p(\mathcal{D})}
\end{equation}
There are many approaches to deal with the posterior distribution in the literature, for example, via Monte Carlo. In this work, we focus on the \emph{variational inference} approach, which optimises a trainable distribution, \(q_{\boldsymbol{\gamma}}(\boldsymbol{\theta})\) to match the true posterior via the parameters, \(\{\boldsymbol{\gamma}\}\). 
In practice, the most common approach is to minimise the Kullback-Leibler (KL) divergence with respect to  $\pi(\boldsymbol{\theta}|\mathcal{D})$. Evaluating the KL divergence and rearranging results in the so-called Evidence Lower Bound (ELBO):
\begin{align*} 
&\text{KL}\left(q_{\boldsymbol{\gamma}}(\boldsymbol{\theta}) | \pi(\boldsymbol{\theta} | \mathcal{D})\right) = \mathbb{E}_{q_{\boldsymbol{\gamma}}(\boldsymbol{\theta})} \left[ \log \frac{q_{\boldsymbol{\gamma}}(\boldsymbol{\theta})}{\pi(\boldsymbol{\theta} | \mathcal{D})} \right] \\ = &\text{KL}(q_{\boldsymbol{\gamma}}(\boldsymbol{\theta}) | p(\boldsymbol{\theta})) + \log p(\mathcal{D}) - \mathbb{E}_{q_{\boldsymbol{\gamma}}(\boldsymbol{\theta})} \left[ \log p(\mathcal{D}| \boldsymbol{\theta}) \right], \\ \implies  &\mathcal{L}_{\boldsymbol{\theta}}(\boldsymbol{\gamma}) := \mathbb{E}_{q_{\boldsymbol{\gamma}}(\boldsymbol{\theta})} \left[ \log p(\mathcal{D}| \boldsymbol{\theta})\right] - \text{KL}(q_{\boldsymbol{\gamma}}(\boldsymbol{\theta}) | p(\boldsymbol{\theta})). 
\end{align*}

Since, $\log p(\mathcal{D})$ is a constant, and the \(\text{KL}\) divergence is non-negative, maximizing the ELBO loss, (\(\mathcal{L}_{\boldsymbol{\theta}}\)), is equivalent to minimizing
$\text{KL}(q_{\boldsymbol{\gamma}}(\theta) \| \pi(\theta | \mathcal{D}))$. \\

\noindent \textbf{Choosing the variational distribution} There are many possibilities one could choose for the variational distribution, \(q_{\boldsymbol{\gamma}}\). However, these choices often come with a tradeoff between expressivity and speed, with even slightly more complicated distributions dealing with the posterior approximation becoming intractable. For initial validation in this work, we choose the simple mean-field Gaussian approximation, and leave more complex choices for future work. The mean-field approach parametrises each of the \(N\) parameters as a normal distribution, where the variational parameters are the mean and variance, \(\{\gamma_i := \{\mu_i, \Sigma^2_i\}\}_{i=1}^N\). Then, we have:
\begin{equation} \label{eqn:mean_field_gaussian_posterior}
    q_{\boldsymbol{\gamma}}(\boldsymbol{\theta}) = \prod_{i=1}^N q^i_{\boldsymbol{\gamma}_i}, \qquad q^i_{\boldsymbol{\gamma}_i} := \mathcal{N}(\theta_i| \mu_i, \Sigma^2_i)
\end{equation}
Assuming we also choose an independent normal distribution for the prior of each parameter, \(p(\theta_i) := \mathcal{N}(\theta_i| 0, 1)\), the ELBO can be written (more) explicitly as:
\begin{equation} \label{eqn:elbo_explicit}
    \mathcal{L}_{\boldsymbol{\theta}}(\boldsymbol{\gamma}) := \underset{q_{\boldsymbol{\gamma}}(\boldsymbol{\theta})}{\mathbb{E}}\left[ \log p(\mathcal{D}| \boldsymbol{\theta})\right] 
    - \frac{1}{2}\sum_{i=1}^N [\Sigma_i^2 + \mu_i^2 -\log(\Sigma_i^2) - 1]. 
\end{equation}
The first term, \(\mathbb{E}_{q_{\boldsymbol{\gamma}}(\boldsymbol{\theta})} \left[ \log p(\mathcal{D}| \boldsymbol{\theta})\right]\), can be estimated via Monte Carlo methods by sampling parameters from the variational distribution.

\subsection{Bayesian quantum machine learning} \label{ssec:bayesian_qml}

Bayesian techniques have appeared in several places in quantum machine learning literature in recent years. Notably, Ref.~\cite{duffield_bayesian_2023} apply Bayesian learning to train parameterised quantum circuits using proximal gradient descent with Laplace priors, which is capable of reducing circuit depth by zeroing out parameters in parameterised operations. They also introduce stochastic gradient Langevin dynamics for Monte Carlo sampling from posterior distributions, highlighting tradeoffs between methods and hyperparameters across several tasks. Extending these techniques to orthogonal quantum networks and comparing them to variational inference remains a promising future direction Ref.~\cite{otterbach_unsupervised_2017} utilises gradient-free Bayesian optimisation with the \texttt{BayesianOptimization} package~\cite{nogueira_bayesian_2014} for Gaussian processes for unsupervised clustering via a quantum approximate optimisation algorithm (QAOA) implemented on Rigetti hardware. Incorporating fault-tolerant quantum primitives into Bayesian methods, Ref.~\cite{zhao_bayesian_2019} propose quantum algorithms for Bayesian training of deep neural networks with Gaussian priors, leveraging quantum states encoding covariance matrices. Ref~\cite{berner_quantum_2021} similarly uses quantum inner product estimation to accelerate Bayesian inference. Finally, Ref.~\cite{benedetti_variational_2021} introduce the quantum circuit Born machine (QCBM) as a variational posterior distribution generator, demonstrating performance advantages over mean-field variational inference methods.

\subsubsection{Metrics} \label{subsubsec:metrics}

The primary metric we use for successful model calibration is the \emph{expected calibration error} (ECE), as follows:
\begin{equation} \label{eqn:ece_metric_definition}
    \text{ECE} := \sum_m \frac{|B_m|}{M} |\text{acc}(B_m)-\text{conf}(B_m)|
\end{equation}

with:
\begin{multline} \label{eqn:acc_confidence_definition}
    \text{acc}(B_m) := \frac{1}{|B_m|}\sum_{i\in B_m}\mathds{1}_{y_i=\hat{y}_i}, \\
    \text{conf}(B_m) := \frac{1}{|B_m|}\sum_{i\in B_m} p_i
\end{multline}

as the accuracy and confidence over each one of \(M\) bins, \(\{B_1, \dots, B_M\}\), with probability of existing in a particular bin as, \(p_i\), with \(i \in B_m, \frac{m-1}{M} < p_i < \frac{m}{M}\). We also have true labels, \(\boldsymbol{y} = \{y_i\}_{i \in B}\), and empirically predicted labels (anomaly or not), \(\widehat{\boldsymbol{y}} = \{\widehat{y}_i\}_{i \in B}\). Intuitively, this metric measures the alignment of the accuracy over a given bin, \(B_m\), with the confidence over that bin. A well-calibrated model is one with low ECE. An undesirable situation is one where the model has high accuracy but low confidence on average in its predictions, in which case the model is unreliable. Conversely, if the accuracy is low, but the model is confident, it may indicate it is being overly conservative.

\section{Bayesian learning of orthogonal neural networks}
\label{Sec:Contribution}

% Now, we are in a position to describe our proposal. We combine the benefits of uncertainty prediction offered by Bayesian methods, with the efficient quantum and quantum-inspired implementations of orthogonal neural networks. 

A na\"{i}ve approach to combining Bayesian learning with orthogonal neural networks might be to apply Bayesian techniques to learn the distribution of the weight matrices directly, followed by an orthogonalisation step, or use SVB~\cite{jia2019orthogonal} techniques. These approaches suffer from the drawbacks discussed in Section~\ref{SubSec:OrthoNNs}, namely, expensive computation or approximation. 

We propose to apply Bayesian techniques instead to the \emph{parameters} (rotation angles) of the quantum circuit gates. The direct connection between the sampled angles and the resulting orthogonal matrix aligns seamlessly with the Bayesian learning paradigm, as it allows for efficient sampling and updates within a probabilistic framework. The method is also scalable, as the cost of applying these rotations scales as $\mathcal{O}(d^2)$ for an $d \times d$ orthogonal matrix.

While this technique can be simulated efficiently on classical computers, it can also be applied as a quantum circuit to execute on quantum devices. This execution on quantum devices can have multi-fold benefits. While the number of gates is quadratic, the depth of the quantum circuit is linear. So, the time to execute this circuit scales linearly on a single quantum computer. Moreover, higher Hamming weight basis states can be used to perform more complex operations in a larger Hilbert space, as discussed in Section~\ref{SubSec:OrthoNNs}.

We assume that each angle is distributed normally, i.e., $\theta \sim \mathcal{N}(\mu_{\theta}, \Sigma_{\theta})$ with some mean $\mu_{\theta}$ and variance $\Sigma_{\theta}$. Our objective is to learn the mean and variance for each angle parameter, which we do using the ELBO method as described in Section~\ref{SubSec:BayesianLearning}. During inference, the rotation parameters are sampled from the learned distribution to apply the corresponding orthogonal layer. We sample the parameters and perform the inference multiple times to generate a distribution of the output of the neural network.

This methodology combines the benefits of Bayesian learning (especially for uncertainty estimation) and quantum/quantum-inspired implementation of orthogonal neural networks for industrial applications in anomaly detection tasks.

\subsection{3D orthogonal convolutions with hamming-weight preserving operations}
\label{SubSec:3D_ortho_conv}

\begin{figure}[h!]
\centering
\begin{tikzpicture}
  \begin{axis}[
    ybar,
    % Make bars wider:
    bar width=18pt,
    % Provide some extra horizontal space so bars don't collide:
    enlarge x limits=0.25,
    % We have 4 x-categories:
    symbolic x coords={FNN, QFNN, 3D-CNN, 3D-QCNN},
    xtick=data,
    xlabel={\sffamily Model},
    ylabel={\sffamily ECE},
    ymin=0.20, ymax=0.27,
    grid=both,
    width=8.5cm,
    height=4.5cm,
    % Place the legend at the top right:
    legend style={
      at={(0.98,0.98)},
      anchor=north east,
      legend columns=1,
      legend cell align=left,
      font=\sffamily
    },
    label style={font=\sffamily},
    ticklabel style={font=\sffamily}
  ]
    \addplot+[
      fill=MyBlue,
      fill opacity=0.9,
      draw=black,
      % Shift bars to the left:
      bar shift=-10pt
    ] coordinates {
      (FNN,0.257)
      (QFNN,0.251)
      (3D-CNN,0.239)
      (3D-QCNN,0.242)
    };
    \addlegendentry{\sffamily\ Point-estimate PE}

    \addplot+[
      fill=MyRed,
      fill opacity=0.9,
      draw=black,
      % Shift bars to the right:
      bar shift=8pt
    ] coordinates {
      (FNN,0.221)
      (QFNN,0.217)
      (3D-CNN,0.209)
      (3D-QCNN,0.230)
    };
    % Add a single legend entry for Bayesian:
    \addlegendentry{\sffamily\ Bayesian learning}

  \end{axis}
\end{tikzpicture}
\caption{\textsf{
\textbf{Model uncertainty measured by ECE.}\\
Bayesian learning versus point-estimate gradient descent for classical and quantum (orthogonal) neural networks within autoencoder anomaly detection pipeline. Bayesian learning outperforms non-Bayesian methods relative to the Estimated Calibration Error (ECE), however orthogonality is less helpful for the more complex model architectures (3D-QCNN).
}
}
\label{fig:ece_evaluation_bayesian_vs_non_bayesian}
\end{figure}

Our first contribution in this area is to describe the new architecture we propose - an orthogonal (quantum) neural network which has the ability to perform three-dimensional convolutions. This extension from 2D operations (as in standard convolutions or quantum vision transformers~\cite{cherrat2022quantum}) is essential to deal with data which has a natural representation in a three-dimensional space. This will be the case for the anomaly detection problem in additive manufacturing that we study. 

\begin{figure}[ht!]
    \centering
    \includegraphics[width=\linewidth]{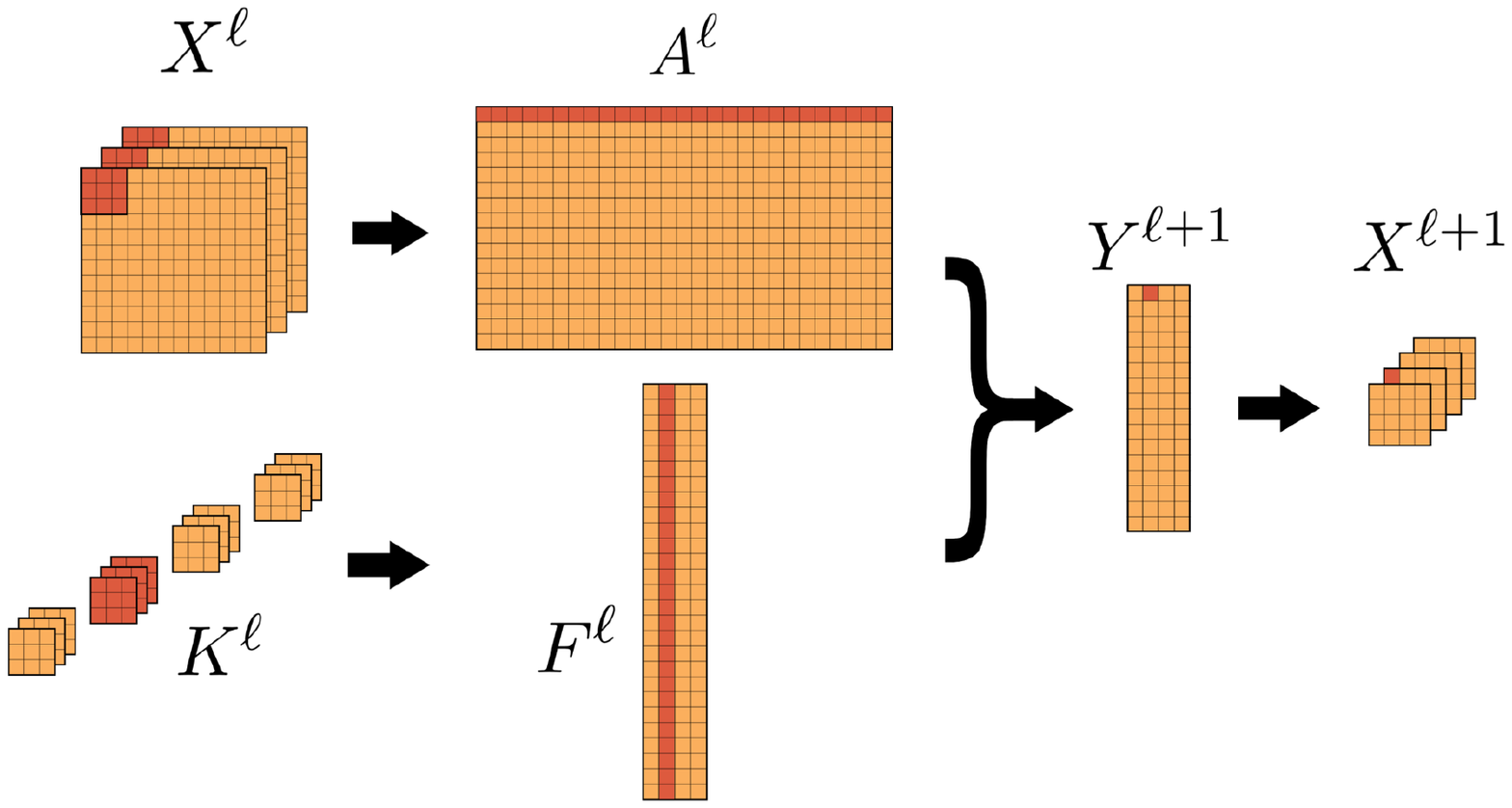}
    \caption{A schematic diagram for showing convolution product reformulated as matrix multiplication for the 2D case (image courtesy \cite{kerenidis2019quantum}). We model the $F^l$ matrix as an orthogonal matrix using quantum circuits.}
    \label{fig:CNNasMM}
\end{figure}

We denote these convolutions as \textsf{OrthoConv3D}, which extend the idea of turning the 2D convolution operation into matrix multiplication (as presented in Ref.~\cite{kerenidis2019quantum}), implemented via orthogonal layers. For the 3D convolutional layer, we flatten each of the $k$ $d \times d \times d$-dimensional convolutional filters - resulting in a matrix, $F^{\ell} \in \mathbb{R}^{k \times d^3}$. Orthogonality is enforced on $F^{\ell}$, so the filters are mutually orthogonal. This reduces the redundancy in the features learned by the kernels. We take each 3D patch of the input object, flatten it, and encode it in the quantum state via a loader. Then, the quantum/quantum-inspired orthogonal layer is applied to multiply the patch with the filters (Figure~\ref{fig:CNNasMM}), and the output is extracted and reshaped accordingly. 
This technique requires $\max(k,d^3)$ qubits to implement, and it does not scale with the size of the data points as is often the case in FNNs.

\section{Results} \label{sec:results}
In this section, we describe the performance of our proposed models on the anomaly detection problem. For all the below results, we use the autoencoder structure as the main anomaly detection architecture as seen in Figure~\ref{fig:overview}. We will either use (orthogonal) feedforward structures in the encoder/decoder, or 3D (orthogonal) convolutions as in Section~\ref{SubSec:3D_ortho_conv}.

\subsection{Anomaly detection details} \label{sub_sec:ad_problem_details}

To test our technique in an industrially relevant setting, we implement it for the use case described in Section~\ref{Sec:UseCase}, the detection of anomalies in additive manufacturing. For this, we utilise a dataset derived from 3D-printed components, with deliberately introduced anomalies designed to facilitate the study of such defect detection algorithms.

The dataset consists of roughly \(5k\) 3D CT scans of objects with each scan being a tensor, \(\boldsymbol{\mathcal{X}} \in \mathbb{R}^{96 \times 96 \times 96}\) in greyscale and normalised to be in the range \([0,1]\). For each data point, we also have a corresponding \emph{mask} of the anomaly, i.e. another \(96 \times 96 \times 96\)-dimensional tensor as follows:
\begin{equation}
    \textsf{Mask}(\boldsymbol{\mathcal{X}})_{ijk} = \begin{cases}
        1 \text{ if voxel } ijk \text{ is anomalous,}\\
        0 \text{ otherwise. }
    \end{cases}
\end{equation}

To process this data for our models, we decompose the scan into \(16 \times 16 \times 16\) voxel blocks and transform the task of masking for the original blocks into a task of classification for the smaller blocks. If a defect exists, i.e. if any voxel set to 1 in the corresponding mask in the small block, we classify it as anomalous; otherwise non-anomalous. This decomposition increases the size of the data \(\times 216\)-fold, so we take only a part of it for training the models.

\begin{figure*}[htbp]
  \centering
  %--- Top Row: ECE, SDA, LuDA ---
  % ECE subplot:
  \begin{subfigure}[b]{0.32\textwidth}
    \centering
    \begin{tikzpicture}
      \begin{axis}[
          ybar,
          xtick={0,1,2,3,4,5,6,7},
          xticklabels={},
          ylabel={\sffamily ECE},
          ymin=2.0, ymax=2.5,
          grid=both,
          minor grid style={dashed,gray!20},
          major grid style={solid,gray!50},
          width=5.5cm,
          height=3cm,
          bar width=12pt,
          every axis plot/.append style={bar shift=0pt}
      ]
        % Bars:
        \addplot[fill=TangoBlue, draw=black, fill opacity=0.4]    coordinates {(0, 2.39)};
        \addplot[fill=TangoCyan, draw=black, fill opacity=0.4]    coordinates {(1, 2.42)};
        \addplot[fill=TangoRed, draw=black, fill opacity=0.4]     coordinates {(2, 2.09)};
        \addplot[fill=TangoOrange, draw=black, fill opacity=0.4]  coordinates {(3, 2.24)};
        \addplot[fill=TangoPurple, draw=black, fill opacity=0.4]   coordinates {(4, 2.15)};
        \addplot[fill=TangoMagenta, draw=black, fill opacity=0.4]  coordinates {(5, 2.29)};
        \addplot[fill=TangoGreen, draw=black, fill opacity=0.4]    coordinates {(6, 2.25)};
        \addplot[fill=TangoButter, draw=black, fill opacity=0.4]   coordinates {(7, 2.31)};
        % Markers:
        \addplot[only marks, mark=square*, mark size=3.5, thick, color=TangoBlue, forget plot]   coordinates {(0, 2.39)};
        \addplot[only marks, mark=triangle*, mark size=3.5, thick, color=TangoCyan, forget plot] coordinates {(1, 2.42)};
        \addplot[only marks, mark=+, mark options={line width=2, mark size=5}, thick, color=TangoRed, forget plot]  coordinates {(2, 2.09)};
        \addplot[only marks, mark=diamond*, mark size=3.5, thick, color=TangoOrange, forget plot] coordinates {(3, 2.24)};
        \addplot[only marks, mark=pentagon*, mark size=3.5, thick, color=TangoPurple, forget plot]   coordinates {(4, 2.15)};
        \addplot[only marks, mark=star, mark size=3.5, thick, color=TangoMagenta, forget plot]       coordinates {(5, 2.29)};
        \addplot[only marks, mark=oplus*, mark size=3.5, thick, color=TangoGreen, forget plot]       coordinates {(6, 2.25)};
        \addplot[only marks, mark=otimes*, mark size=3.5, thick, color=TangoButter, forget plot]      coordinates {(7, 2.31)};
      \end{axis}
    % \node[anchor=south] at ($(current bounding box.north)+(-0.5,0)$) {\(\times 10^{-1}\)};
    \end{tikzpicture}
  \vspace{-0.4cm}
    \caption{ECE (\(\times 10^{-1}\)) (\(\boldsymbol{\downarrow}\))}
  \end{subfigure}
  % SDA subplot:
  \begin{subfigure}[b]{0.32\textwidth}
    \centering
    \begin{tikzpicture}
      \begin{axis}[
          ybar,
          xtick={0,1,2,3,4,5,6,7},
          xticklabels={},
          ylabel={\sffamily SDA},
          ymin=3.5, ymax=4.5,
          grid=both,
          minor grid style={dashed,gray!20},
          major grid style={solid,gray!50},
          width=5.5cm,
          height=3cm,
          bar width=12pt,
          every axis plot/.append style={bar shift=0pt}
      ]
        % All SDA values are 4.
        \addplot[fill=TangoBlue, draw=black, fill opacity=0.4]    coordinates {(0,4)};
        \addplot[fill=TangoCyan, draw=black, fill opacity=0.4]    coordinates {(1,4)};
        \addplot[fill=TangoRed, draw=black, fill opacity=0.4]     coordinates {(2,4)};
        \addplot[fill=TangoOrange, draw=black, fill opacity=0.4]  coordinates {(3,4)};
        \addplot[fill=TangoPurple, draw=black, fill opacity=0.4]   coordinates {(4,4)};
        \addplot[fill=TangoMagenta, draw=black, fill opacity=0.4]  coordinates {(5,4)};
        \addplot[fill=TangoGreen, draw=black, fill opacity=0.4]    coordinates {(6,4)};
        \addplot[fill=TangoButter, draw=black, fill opacity=0.4]   coordinates {(7,4)};
        % Markers:
        \addplot[only marks, mark=square*, mark size=3.5, thick, color=TangoBlue, forget plot]   coordinates {(0,4)};
        \addplot[only marks, mark=triangle*, mark size=3.5, thick, color=TangoCyan, forget plot] coordinates {(1,4)};
        \addplot[only marks, mark=+, mark options={line width=2, mark size=5}, thick, color=TangoRed, forget plot]  coordinates {(2,4)};
        \addplot[only marks, mark=diamond*, mark size=3.5, thick, color=TangoOrange, forget plot] coordinates {(3,4)};
        \addplot[only marks, mark=pentagon*, mark size=3.5, thick, color=TangoPurple, forget plot]   coordinates {(4,4)};
        \addplot[only marks, mark=star, mark size=3.5, thick, color=TangoMagenta, forget plot]       coordinates {(5,4)};
        \addplot[only marks, mark=oplus*, mark size=3.5, thick, color=TangoGreen, forget plot]       coordinates {(6,4)};
        \addplot[only marks, mark=otimes*, mark size=3.5, thick, color=TangoButter, forget plot]      coordinates {(7,4)};
      \end{axis}
    \end{tikzpicture}
      \vspace{-0.4cm}
    \caption{SDA}
  \end{subfigure}
  % LuDA subplot:
  \begin{subfigure}[b]{0.32\textwidth}
    \centering
    \begin{tikzpicture}
      \begin{axis}[
          ybar,
          xtick={0,1,2,3,4,5,6,7},
          xticklabels={},
          ylabel={\sffamily LuDA},
          ymin=12, ymax=18,
          grid=both,
          minor grid style={dashed,gray!20},
          major grid style={solid,gray!50},
          width=5.5cm,
          height=3cm,
          bar width=12pt,
          every axis plot/.append style={bar shift=0pt}
      ]
        \addplot[fill=TangoBlue, draw=black, fill opacity=0.4]    coordinates {(0,14)};
        \addplot[fill=TangoCyan, draw=black, fill opacity=0.4]    coordinates {(1,15)};
        \addplot[fill=TangoRed, draw=black, fill opacity=0.4]     coordinates {(2,14)};
        \addplot[fill=TangoOrange, draw=black, fill opacity=0.4]  coordinates {(3,17)};
        \addplot[fill=TangoPurple, draw=black, fill opacity=0.4]   coordinates {(4,14)};
        \addplot[fill=TangoMagenta, draw=black, fill opacity=0.4]  coordinates {(5,15)};
        \addplot[fill=TangoGreen, draw=black, fill opacity=0.4]    coordinates {(6,14)};
        \addplot[fill=TangoButter, draw=black, fill opacity=0.4]   coordinates {(7,15)};
        % Markers:
        \addplot[only marks, mark=square*, mark size=3.5, thick, color=TangoBlue, forget plot]   coordinates {(0,14)};
        \addplot[only marks, mark=triangle*, mark size=3.5, thick, color=TangoCyan, forget plot] coordinates {(1,15)};
        \addplot[only marks, mark=+, mark options={line width=2, mark size=5}, thick, color=TangoRed, forget plot]  coordinates {(2,14)};
        \addplot[only marks, mark=diamond*, mark size=3.5, thick, color=TangoOrange, forget plot] coordinates {(3,17)};
        \addplot[only marks, mark=pentagon*, mark size=3.5, thick, color=TangoPurple, forget plot]   coordinates {(4,14)};
        \addplot[only marks, mark=star, mark size=3.5, thick, color=TangoMagenta, forget plot]       coordinates {(5,15)};
        \addplot[only marks, mark=oplus*, mark size=3.5, thick, color=TangoGreen, forget plot]       coordinates {(6,14)};
        \addplot[only marks, mark=otimes*, mark size=3.5, thick, color=TangoButter, forget plot]      coordinates {(7,15)};
      \end{axis}
    \end{tikzpicture}
  \vspace{-0.4cm}
    \caption{LuDA}
  \end{subfigure}\\ 
  \vspace{0.2cm}
  %--- Bottom Row: Precision, Recall, F1-Score ---
  % Precision subplot:
  \begin{subfigure}[b]{0.32\textwidth}
    \centering
    \begin{tikzpicture}
      \begin{axis}[
          ybar,
          xtick={0,1,2,3,4,5,6,7},
          xticklabels={},
          ylabel={\sffamily Precision},
          ymin=72, ymax=78,
          grid=both,
          minor grid style={dashed,gray!20},
          major grid style={solid,gray!50},
          width=5.5cm,
          height=3cm,
          bar width=12pt,
          every axis plot/.append style={bar shift=0pt}
      ]
         \addplot[fill=TangoBlue, draw=black, fill opacity=0.4]    coordinates {(0,76.35)};
         \addplot[fill=TangoCyan, draw=black, fill opacity=0.4]    coordinates {(1,75.49)};
         \addplot[fill=TangoRed, draw=black, fill opacity=0.4]     coordinates {(2,73.98)};
         \addplot[fill=TangoOrange, draw=black, fill opacity=0.4]  coordinates {(3,73.13)};
         \addplot[fill=TangoPurple, draw=black, fill opacity=0.4]   coordinates {(4,76.77)};
         \addplot[fill=TangoMagenta, draw=black, fill opacity=0.4]  coordinates {(5,75.65)};
         \addplot[fill=TangoGreen, draw=black, fill opacity=0.4]    coordinates {(6,76.95)};
         \addplot[fill=TangoButter, draw=black, fill opacity=0.4]   coordinates {(7,75.95)};
         % Markers:
         \addplot[only marks, mark=square*, mark size=3.5, thick, color=TangoBlue, forget plot]   coordinates {(0,76.35)};
         \addplot[only marks, mark=triangle*, mark size=3.5, thick, color=TangoCyan, forget plot] coordinates {(1,75.49)};
         \addplot[only marks, mark=+, mark options={line width=2, mark size=5}, thick, color=TangoRed, forget plot]  coordinates {(2,73.98)};
         \addplot[only marks, mark=diamond*, mark size=3.5, thick, color=TangoOrange, forget plot] coordinates {(3,73.13)};
         \addplot[only marks, mark=pentagon*, mark size=3.5, thick, color=TangoPurple, forget plot]   coordinates {(4,76.77)};
         \addplot[only marks, mark=star, mark size=3.5, thick, color=TangoMagenta, forget plot]       coordinates {(5,75.65)};
         \addplot[only marks, mark=oplus*, mark size=3.5, thick, color=TangoGreen, forget plot]       coordinates {(6,76.95)};
         \addplot[only marks, mark=otimes*, mark size=3.5, thick, color=TangoButter, forget plot]      coordinates {(7,75.95)};
      \end{axis}
    \end{tikzpicture}
      \vspace{-0.4cm}
    \caption{Precision (\(\boldsymbol{\uparrow}\))}
  \end{subfigure}
  % Recall subplot:
  \begin{subfigure}[b]{0.32\textwidth}
    \centering
    \begin{tikzpicture}
      \begin{axis}[
          ybar,
          xtick={0,1,2,3,4,5,6,7},
          xticklabels={},
          ylabel={\sffamily Recall},
          ymin=72, ymax=76,
          grid=both,
          minor grid style={dashed,gray!20},
          major grid style={solid,gray!50},
          width=5.5cm,
          height=3cm,
          bar width=12pt,
          every axis plot/.append style={bar shift=0pt}
      ]
         \addplot[fill=TangoBlue, draw=black, fill opacity=0.4]    coordinates {(0,74.97)};
         \addplot[fill=TangoCyan, draw=black, fill opacity=0.4]    coordinates {(1,74.12)};
         \addplot[fill=TangoRed, draw=black, fill opacity=0.4]     coordinates {(2,72.86)};
         \addplot[fill=TangoOrange, draw=black, fill opacity=0.4]  coordinates {(3,72.37)};
         \addplot[fill=TangoPurple, draw=black, fill opacity=0.4]   coordinates {(4,74.78)};
         \addplot[fill=TangoMagenta, draw=black, fill opacity=0.4]  coordinates {(5,74.43)};
         \addplot[fill=TangoGreen, draw=black, fill opacity=0.4]    coordinates {(6,75.12)};
         \addplot[fill=TangoButter, draw=black, fill opacity=0.4]   coordinates {(7,74.22)};
         % Markers:
         \addplot[only marks, mark=square*, mark size=3.5, thick, color=TangoBlue, forget plot]   coordinates {(0,74.97)};
         \addplot[only marks, mark=triangle*, mark size=3.5, thick, color=TangoCyan, forget plot] coordinates {(1,74.12)};
         \addplot[only marks, mark=+, mark options={line width=2, mark size=5}, thick, color=TangoRed, forget plot]  coordinates {(2,72.86)};
         \addplot[only marks, mark=diamond*, mark size=3.5, thick, color=TangoOrange, forget plot] coordinates {(3,72.37)};
         \addplot[only marks, mark=pentagon*, mark size=3.5, thick, color=TangoPurple, forget plot]   coordinates {(4,74.78)};
         \addplot[only marks, mark=star, mark size=3.5, thick, color=TangoMagenta, forget plot]       coordinates {(5,74.43)};
         \addplot[only marks, mark=oplus*, mark size=3.5, thick, color=TangoGreen, forget plot]       coordinates {(6,75.12)};
         \addplot[only marks, mark=otimes*, mark size=3.5, thick, color=TangoButter, forget plot]      coordinates {(7,74.22)};
      \end{axis}
    \end{tikzpicture}
      \vspace{-0.4cm}
    \caption{Recall (\(\boldsymbol{\uparrow}\))}
  \end{subfigure}
  % F1-Score subplot:
  \begin{subfigure}[b]{0.32\textwidth}
    \centering
    \begin{tikzpicture}
      \begin{axis}[
          ybar,
          xtick={0,1,2,3,4,5,6,7},
          xticklabels={},
          ylabel={\sffamily F1-Score},
          ymin=72, ymax=77,
          grid=both,
          minor grid style={dashed,gray!20},
          major grid style={solid,gray!50},
          width=5.5cm,
          height=3cm,
          bar width=12pt,
          every axis plot/.append style={bar shift=0pt}
      ]
         \addplot[fill=TangoBlue, draw=black, fill opacity=0.4]    coordinates {(0,75.65)};
         \addplot[fill=TangoCyan, draw=black, fill opacity=0.4]    coordinates {(1,74.79)};
         \addplot[fill=TangoRed, draw=black, fill opacity=0.4]     coordinates {(2,73.41)};
         \addplot[fill=TangoOrange, draw=black, fill opacity=0.4]  coordinates {(3,72.74)};
         \addplot[fill=TangoPurple, draw=black, fill opacity=0.4]   coordinates {(4,75.76)};
         \addplot[fill=TangoMagenta, draw=black, fill opacity=0.4]  coordinates {(5,75.03)};
         \addplot[fill=TangoGreen, draw=black, fill opacity=0.4]    coordinates {(6,76.02)};
         \addplot[fill=TangoButter, draw=black, fill opacity=0.4]   coordinates {(7,75.07)};
         % Markers:
         \addplot[only marks, mark=square*, mark size=3.5, thick, color=TangoBlue, forget plot]   coordinates {(0,75.65)};
         \addplot[only marks, mark=triangle*, mark size=3.5, thick, color=TangoCyan, forget plot] coordinates {(1,74.79)};
         \addplot[only marks, mark=+, mark options={line width=2, mark size=5}, thick, color=TangoRed, forget plot]  coordinates {(2,73.41)};
         \addplot[only marks, mark=diamond*, mark size=3.5, thick, color=TangoOrange, forget plot] coordinates {(3,72.74)};
         \addplot[only marks, mark=pentagon*, mark size=3.5, thick, color=TangoPurple, forget plot]   coordinates {(4,75.76)};
         \addplot[only marks, mark=star, mark size=3.5, thick, color=TangoMagenta, forget plot]       coordinates {(5,75.03)};
         \addplot[only marks, mark=oplus*, mark size=3.5, thick, color=TangoGreen, forget plot]       coordinates {(6,76.02)};
         \addplot[only marks, mark=otimes*, mark size=3.5, thick, color=TangoButter, forget plot]      coordinates {(7,75.07)};
      \end{axis}
    \end{tikzpicture}
      \vspace{-0.4cm}
    \caption{F1-Score (\(\boldsymbol{\uparrow}\))}
  \end{subfigure}
  % Legend:
  \begin{subfigure}[b]{\textwidth}
    \centering
    \captionsetup{labelformat=empty}
    \begin{tikzpicture}[baseline=(current bounding box.center)]
      \matrix (m) [matrix of nodes, 
                   nodes={anchor=west, align=center},
                   column sep=0.01cm, row sep=0.05cm]{
        % First row: PE and Bayesian entries
        {\tikz \draw[mark=square*, mark size=5, color=TangoBlue] plot coordinates {(0,0)};} & {\sffamily\footnotesize 3D-CNN (PE)} & & &
        {\tikz \draw[mark=triangle*, mark size=5, color=TangoCyan] plot coordinates {(0,0)};} & {\sffamily\footnotesize 3D-QCNN (PE)} & & &
        {\tikz \draw[mark=+, mark options={line width=2, mark size=5}, color=TangoRed] plot coordinates {(0,0)};} & {\sffamily\footnotesize 3D-CNN (Bayesian)} & &&
        {\tikz \draw[mark=diamond*, mark size=5, color=TangoOrange] plot coordinates {(0,0)};} & {\sffamily\footnotesize 3D-QCNN (Bayesian)} \\
        % Second row: MCD and Ensemble entries
        {\tikz \draw[mark=pentagon*, mark size=5, color=TangoPurple] plot coordinates {(0,0)};} & {\sffamily\footnotesize 3D-CNN (MCD)} & & &
        {\tikz \draw[mark=star, mark options={line width=2, mark size=5}, color=TangoMagenta] plot coordinates {(0,0)};} & {\sffamily\footnotesize 3D-QCNN (MCD)} & & &
        {\tikz \draw[mark=oplus*, mark size=5, color=TangoGreen] plot coordinates {(0,0)};} & {\sffamily\footnotesize 3D-CNN (Ensemble)} & & & 
        {\tikz \draw[mark=otimes*, mark size=5, color=TangoButter] plot coordinates {(0,0)};} & {\sffamily\footnotesize 3D-QCNN (Ensemble)} \\
      };
    \end{tikzpicture}
  \end{subfigure}
  \caption{\small
  \textsf{
  \textbf{Comparing uncertainty prediction approaches with 3D architectures}.
  Each plot shows a different metric for each approach. a) Estimated Calibration Error (ECE), our quantifier for robust predictions. b) The smallest undetected anomaly (SDA) and largest undetected anomaly (LuDA) for each model. Finally, standard metrics: c) Precision, d) Recall and e) F1-Score. For both 3D-CNN and 3D-QCNN architectures, we test point-estimate gradient descent (PE), Bayesian learning, Monte Carlo Dropout (MCD) and Ensembling. Focusing on ECE again, Bayesian learning is the superior uncertainty prediction model, however, for this architecture, orthogonal components are not as helpful.
  }
  }
  \label{fig:anomaly_results_new}
\end{figure*}

\subsection{Methodology}
\label{sub_sec:methodology}
We use the standard approach of using autoencoders for anomaly detection as motivated in Section \ref{subsec:ae_for_ad}. The autoencoder model takes the 3D representation of the object as voxels, \(\boldsymbol{\mathcal{X}} \in \mathbb{R}^{d_1 \times d_1 \times d_1}\) and compresses to a latent space via the \emph{encoder}, \(E(\boldsymbol{\mathcal{X}}) = \boldsymbol{\mathcal{Y}} \in \mathbb{R}^{d_2 \times d_2 \times d_2}\). A \emph{decoder} then decodes it from the latent space to recreate the original object, \(D(\boldsymbol{\mathcal{Y}}) = \boldsymbol{\mathcal{Z}} \in \mathbb{R}^{d_1 \times d_1 \times d_1}\). If the autoencoder is capable of re-constructing the image with low reconstruction error, in other words, \(D(E(\boldsymbol{\mathcal{X}}))\approx \boldsymbol{\mathcal{X}}\), we consider it non-anomalous. Otherwise, we classify it as an anomaly. The intuition is that anomalies act as outliers whose features cannot be easily captured by the model, as they will be significantly different from the surrounding features. Here, we test different model families for the encoder and decoder neural network layers as described below. For all the below results, we use the \emph{butterfly}~\cite{cherrat2022quantum} layout for the parameterisation of orthogonal layers. 
% The butterfly layer is most suitable for quantum hardware which has qubits natively connected in an all-to-all manner, due to the non-local operation of the underlying \(RBS\) gates. Moreover, it has the least possible depth (logarithmic in the number of qubits) for a fully-connected layer.

For the below results, we also compare to two other uncertainty quantification methods, Monte Carlo Dropout (MCD)~\cite{gal_dropout_2016} and Ensembling~\cite{lakshminarayanan_simple_2017}. Monte Carlo Dropout involves including a dropout~\cite{srivastava_dropout_2014} layer between the encoder and decoder, and also just before the final output of the autoencoder. It was shown in Ref.~\cite{gal_dropout_2016} that such dropout approximately creates a Bayesian approximation to a deep Gaussian process in a network and hence can be used for uncertainty quantification.

On the other hand, one approach to applying Ensemble methods for uncertainty prediction~\cite{lakshminarayanan_simple_2017} uses ensembles of predictor models and averages the prediction of each individual network, to produce the final result. We take this same approach in this work - for each of the model families below, the output of the ensemble is simply taken to be the average of each sub-network.

\begin{figure} [h!]
    \centering
    \includegraphics[width=\columnwidth]{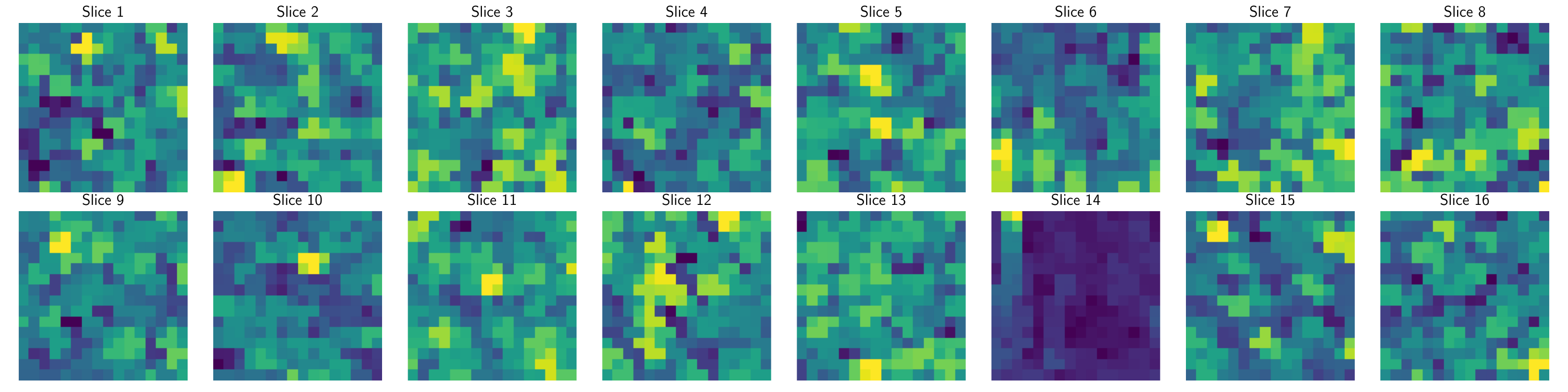}
    \caption{
    \textsf{
    \textbf{Decomposed anomalous voxel.} \\
    \(16\) slices of \(16 \times 16\) pixels. We choose slice \(14\) containing an anomaly to focus on for the hardware experiments.
    }
    }
    \label{fig:flattened_anomaly_voxels}
\end{figure}

\begin{figure*}[ht!]
    \centering
    \begin{subfigure}{0.49\columnwidth}
        \centering
        \begin{tikzpicture}
            \begin{axis}[
                xlabel={\sffamily Indices},
                ylabel={\sffamily Fid. (rand)},
                ymin=0.9, ymax=1.0,
                xtick={1,2,3,4,5,6,7,8},
                xticklabel style={font=\sffamily\small},
                yticklabel style={font=\sffamily\small},
                grid=both,
                minor grid style={dashed,gray!20},
                major grid style={solid,gray!50},
                width=4cm,
                height=3.5cm,
                legend to name=legendhwone, 
                legend columns=2,  
                legend cell align = left,
                % transpose legend = true,
                legend style={
                    font=\sffamily\small,
                    align=left,
                    fill=white,
                    draw=none
                }
            ]

            \addplot[mark=square*,  
                    mark size=3, 
                    thick, smooth, color={rgb,255:red,255; green,127; blue,14}]
                coordinates {
                    (1,0.9993181) (2,0.9986265) (3,0.99771166) (4,0.9991903)
                    (5,0.9976276) (6,0.9963859) (7,0.99938214) (8,0.9990711)
                };
            \addlegendentry{QASM (sim) \(M=1\text{k}\)}
            
            \addplot[mark=triangle*,  
                    mark size=3, 
                     thick, smooth, color={rgb,255:red,44; green,160; blue,44}]
                coordinates {
                    (1,0.99987924) (2,0.9998306) (3,0.9998995) (4,0.9998747)
                    (5,0.99978805) (6,0.9998989) (7,0.99992144) (8,0.99980617)
                };
            \addlegendentry{QASM (sim) \(M=10\text{k}\)}

            \addplot[mark=*,  
                    mark size=3, 
                    thick, smooth, color={rgb,255:red,31; green,119; blue,180}]
                coordinates {
                    (1,0.9996598) (2,0.999451) (3,0.9997399) (4,0.9995788)
                    (5,0.9995647) (6,0.99984896) (7,0.99981785) (8,0.9993472)
                };
            \addlegendentry{Fake Brisbane (sim) \(M=10\text{k}\)}
            
            \addplot[mark=diamond*, 
                    mark size=3,
            thick, smooth, color={rgb,255:red,214; green,39; blue,40}]
                coordinates {
                    (1,0.97106194) (2,0.95999503) (3,0.9576139) (4,0.94477254)
                    (5,0.91764504) (6,0.9810489) (7,0.9806591) (8,0.92345905)
                };
            \addlegendentry{Brisbane (QPU) \(M=1\text{k}\)}
            
            \addplot[mark=x, 
                    mark options={line width=3, mark size=3},
                    thick, smooth, color={rgb,255:red,148; green,103; blue,189}]
                coordinates {
                    (1,0.97868973) (2,0.9792321) (3,0.9577398) (4,0.9541953)
                    (5,0.915) (6,0.9617446) (7,0.9737799) (8,0.93597525)
                };
            \addlegendentry{Brisbane (QPU) \(M=10\text{k}\)}
            
            \end{axis}
        \end{tikzpicture}
        \subcaption{}
    \end{subfigure}
    \begin{subfigure}{0.49\columnwidth}
        \centering
\begin{tikzpicture}
    \begin{axis}[
        ybar,
        % Place ticks at x=0..4
        xtick={0,1,2,3,4},
        % Label each tick with the category name
        xticklabels={
            {},
            {},
            {},
            {},
            {}
        },
        ymin=0.9, ymax=1.0,
        grid=both,
        minor grid style={dashed,gray!20},
        major grid style={solid,gray!50},
        width=4cm,
        height=3.5cm,
        bar width=10pt,
        % markersize=3,
        xticklabel style={font=\sffamily\small, rotate=90, anchor=east},
        xlabel={\sffamily Device},
        ylabel={\sffamily Avg. fid. (rand)},
        % Disable any auto-shifting/grouping of bars
        cycle list={{}}
    ]
    
    %--- One bar per \addplot, each at x=0..4 with bar shift=0pt
    
    \addplot[
        fill={rgb,255:red,255; green,127; blue,14}, 
        draw=black,
              fill opacity=0.4,
        bar shift=0pt
    ] coordinates {(0,0.99841416)};
    
    \addplot[
        fill={rgb,255:red,44; green,160; blue,44}, 
        draw=black,
      fill opacity=0.4,
        bar shift=0pt
    ] coordinates {(1,0.9998623)};

     \addplot[
        fill={rgb,255:red,31; green,119; blue,180}, 
        draw=black,
      fill opacity=0.4,
        bar shift=0pt
    ] coordinates {(2,0.99962604)};
    
    \addplot[
        fill={rgb,255:red,214; green,39; blue,40}, 
        draw=black,
      fill opacity=0.4,
        bar shift=0pt
    ] coordinates {(3,0.9545319)};
    
    \addplot[
        fill={rgb,255:red,148; green,103; blue,189}, 
        draw=black,
          fill opacity=0.4,
        bar shift=0pt
    ] coordinates {(4,0.9570446)};
    
    %--- Markers at the top of each bar (same x coords, same colors)
    
    \addplot[
        only marks,
        mark=square*,
        mark size=4.5,
        thick,
        color={rgb,255:red,255; green,127; blue,14},
        forget plot
    ] coordinates {(0,0.99841416)};
    
    \addplot[
        only marks,
        mark=triangle*,
        mark size=4.5,
        thick,
        color={rgb,255:red,44; green,160; blue,44},
        forget plot
    ] coordinates {(1,0.9998623)};

    \addplot[
        only marks,
        mark=*,
        mark size=5,
        thick,
        color={rgb,255:red,31; green,119; blue,180},
        forget plot
    ] coordinates {(2,0.99962604)};
    
    \addplot[
        only marks,
        mark=diamond*,
        mark size=4.5,
        thick,
        color={rgb,255:red,214; green,39; blue,40},
        forget plot
    ] coordinates {(3,0.9545319)};
    
    \addplot[
        only marks,
        mark=x,
        mark options={line width=3, mark size=4.5},
        thick,
        color={rgb,255:red,148; green,103; blue,189},
        forget plot
    ] coordinates {(4,0.9570446)};
    
    \end{axis}
\end{tikzpicture}
        \subcaption{}
    \end{subfigure}
     \begin{subfigure}{0.49\columnwidth}
        \centering
        \begin{tikzpicture}
            \begin{axis}[
                xlabel={\sffamily Indices},
                ylabel={\sffamily Fidelity (voxel)},
                ymin=0.90, ymax=1.0,
                xtick={0, 1,2,3,4,5,6,7},
                xticklabel style={font=\sffamily\small},
                yticklabel style={font=\sffamily\small},
                grid=both,
                minor grid style={dashed,gray!20},
                major grid style={solid,gray!50},
                width=4cm,
                height=3.5cm,
                legend to name=legendhwtwo, 
                legend columns=3,  
                legend cell align = left,
                % transpose legend = true,
                legend style={
                    font=\sffamily\small,
                    align=left,
                    fill=white,
                    draw=none
                }
            ]
            
            \addplot[mark=square*,  
                    mark size=3, 
                    thick, smooth, color={rgb,255:red,255; green,127; blue,14}]
                coordinates {
                    (0 ,0.99937737 ) (1, 0.99666065) (2, 0.99828863) (3, 0.9982118)
                    (4, 0.9981998) (5, 0.9978008) (6, 0.99878263) (7,  0.99830294)
                };
            \addlegendentry{QASM (sim) \(M=1\text{k}\)}
            
            \addplot[mark=triangle*,  
                    mark size=3, 
                     thick, smooth, color={rgb,255:red,44; green,160; blue,44}]
                coordinates {
                    (0, 0.9996828)  (1, 0.9997591)  (2, 0.9995373)  (3, (0.9999465)  (4, 0.99962)    (5, 0.99984837)
                    (6, 0.9996809)  (7, 0.9997299) 
                };
            \addlegendentry{QASM (sim) \(M=10\text{k}\)}

            % Data & Colors (Tableau-inspired)
            \addplot[mark=*,  
                    mark size=3, 
                    thick, smooth, color={rgb,255:red,31; green,119; blue,180}]
                coordinates {
                    (0, 0.9995548)  (1, 0.9994561) (2,  0.9995988) (3,  0.99958307) (4, 0.99984825) (5, 0.99946886)
                    (6, 0.9997337)  (7, 0.999188)  
                };
            \addlegendentry{Fake Brisbane (sim) \(M=10\text{k}\)}
            
            \addplot[mark=diamond*, 
                    mark size=3,
            thick, smooth, color={rgb,255:red,214; green,39; blue,40}]
                coordinates {
                    (0, 0.93861955) (1, 0.9334466)  (2, 0.96413434) (3, 0.9489824)  (4, 0.97094226) (5, 0.9607001)
                    (6, 0.9751152)  (7, 0.98674655)
                };
            \addlegendentry{Brisbane (QPU) \(M=1\text{k}\)}
            
            \addplot[mark=x, 
                    mark options={line width=3, mark size=3},
                    thick, smooth, color={rgb,255:red,148; green,103; blue,189}]
                coordinates {
                    (0, 0.9814528) (1,  0.9633939)  (2, 0.9726879) (3, 0.9628842)  (4, 0.99428624)(5, 0.9881039)
 (6, 0.98973966) (7, 0.98187405)
                };
            \addlegendentry{Brisbane (QPU) \(M=10\text{k}\)}
            
            \end{axis}
        \end{tikzpicture}
        \subcaption{}
    \end{subfigure}
    \begin{subfigure}{0.49\columnwidth}
        \centering
\begin{tikzpicture}
    \begin{axis}[
        ybar,
        % Place ticks at x=0..4
        xtick={0,1,2,3,4},
        % Label each tick with the category name
        xticklabels={
            {},
            {},
            {},
            {},
            {}
        },
        ymin=0.90, ymax=1.0,
        grid=both,
        minor grid style={dashed,gray!20},
        major grid style={solid,gray!50},
        width=4cm,
        height=3.5cm,
        bar width=10pt,
        % markersize=3,
        xticklabel style={font=\sffamily\small, rotate=90, anchor=east},
        xlabel={\sffamily Device},
        ylabel={\sffamily Avg. fid. (voxel)},
        % Disable any auto-shifting/grouping of bars
        cycle list={{}}
    ]
    
    %--- One bar per \addplot, each at x=0..4 with bar shift=0pt
    
    \addplot[
        fill={rgb,255:red,255; green,127; blue,14}, 
        draw=black,
              fill opacity=0.4,
        bar shift=0pt
    ] coordinates {(0, 0.9982031)};
    
    \addplot[
        fill={rgb,255:red,44; green,160; blue,44}, 
        draw=black,
      fill opacity=0.4,
        bar shift=0pt
    ] coordinates {(1, 0.99972564)};

     \addplot[
        fill={rgb,255:red,31; green,119; blue,180}, 
        draw=black,
      fill opacity=0.4,
        bar shift=0pt
    ] coordinates {(2, 0.9995539)};
    
    \addplot[
        fill={rgb,255:red,214; green,39; blue,40}, 
        draw=black,
      fill opacity=0.4,
        bar shift=0pt
    ] coordinates {(3, 0.9598359)};
    
    \addplot[
        fill={rgb,255:red,148; green,103; blue,189}, 
        draw=black,
          fill opacity=0.4,
        bar shift=0pt
    ] coordinates {(4,0.9793028)};
    
    %--- Markers at the top of each bar (same x coords, same colors)
    
    \addplot[
        only marks,
        mark=square*,
        mark size=4.5,
        thick,
        color={rgb,255:red,255; green,127; blue,14},
        forget plot
    ] coordinates {(0, 0.9982031)};
    
    \addplot[
        only marks,
        mark=triangle*,
        mark size=4.5,
        thick,
        color={rgb,255:red,44; green,160; blue,44},
        forget plot
    ] coordinates {(1, 0.99972564)};

    \addplot[
        only marks,
        mark=*,
        mark size=5,
        thick,
        color={rgb,255:red,31; green,119; blue,180},
        forget plot
    ] coordinates {(2, 0.9995539)};
    
    \addplot[
        only marks,
        mark=diamond*,
        mark size=4.5,
        thick,
        color={rgb,255:red,214; green,39; blue,40},
        forget plot
    ] coordinates {(3, 0.9598359)};
    
    \addplot[
        only marks,
        mark=x,
        mark options={line width=3, mark size=4.5},
        thick,
        color={rgb,255:red,148; green,103; blue,189},
        forget plot
    ] coordinates {(4,0.9793028)};
    
    \end{axis}
\end{tikzpicture}

        \subcaption{}
    \end{subfigure}
    \begin{center}
        % This pulls in the legend from the "legend to name=commonlegend" above
        \pgfplotslegendfromname{legendhwone}
    \end{center}
    \caption{
    \textsf{
    \textbf{Fidelity plots for ortholinear circuits.}\\
    Fidelity plots incorporating only shot noise (QASM simulator), simulated device noise for IBM Brisbane QPU (`Fake' Brisbane) and the real Brisbane QPU, using \(8\) qubits in all cases. Number of shots taken to measure fidelity in each case is \(M \in \{1\text{k}, 10\text{k}\}\). (a) Fidelity for each of the \(8\) input vectors, (b) average fidelity across all inputs. Input vectors to the ortholinear circuit are randomly chosen \textbf{(a, b)} are downscaled vectors which compose the voxel slice \(14\) from Fig.~\ref{fig:flattened_anomaly_voxels} \textbf{(c, d)}.
    }}
    \label{fig:hardware_fidelity_experiment}
\end{figure*}

\subsection{Fully connected (orthogonal) neural networks} \label{sub_sec:fnn_results}

As our baseline, we begin with \emph{fully connected} layers within the anomaly detection autoencoder. We test both standard linear layers, \(\boldsymbol{y} = \mathbf{W}\boldsymbol{x}\) and \emph{orthogonal} linear layers \(\boldsymbol{y} = \mathbf{O}_{U(\boldsymbol{\theta})}\boldsymbol{x}\) where \(\mathbf{O}_{U(\boldsymbol{\theta)}}\) is an orthogonal matrix induced by the Hamming-weight preserving unitaries, \(U(\boldsymbol{\theta)}\) as described above. The encoder comprises two linear layers $\mathbf{W}_1, \mathbf{W}_2$, which are $128$ and \(64\)-dimensional (mapping to a latent space)  respectively with a \texttt{ReLU} activation between. The decoder includes a \texttt{ReLU} activation, a $128$-dimensional linear layer $\mathbf{W}_3$ with \texttt{tanh} activation, and a final linear layer $\mathbf{W}_4$, whose output is reshaped to match the original $3$-dimensional tensor. The ``quantum'' variant replaces internal layers $\mathbf{W}_2, \mathbf{W}_3$ with orthogonal versions $\mathbf{O}_{U_2(\boldsymbol{\theta})}, \mathbf{O}_{U_3(\boldsymbol{\theta})}$, leaving other aspects unchanged. We denote the model with purely classical layers as \emph{feedforward} neural network (FNN), and the model with hybrid quantum-classical layers as quantum FNN (QFNN).

First, Table~\ref{tab:fnn_results} shows the effectiveness of Bayesian learning on both of these model families for anomaly detection. We see that for both classical (FNN) and quantum (QFNN) models, Bayesian learning approaches outperform their standard gradient descent-trained point-estimate (PE) counterparts, when the Estimated Calibration Error (ECE) is the metric of interest. When standard classification metrics (precision, recall \& F1 score) are the baseline, however, the non-Bayesian approach is superior. This demonstrates that when one requires a \emph{robust} model with uncertainty quantification, Bayesian training is beneficial. Also, note that the orthogonal versions outperform their non-orthogonal counterparts with slightly fewer parameters - the QFNN model used \(\sim 1.4\text{M}\) parameters compared to  \(\sim 1.5\text{M}\) for the FNN. In the tables, we also show the smallest detected anomaly (SDA) and the largest undetected anomaly (LuDA). The SDA/LuDA is the minimum/maximum size of the anomalous pixels in the data point, among all correctly/incorrectly identified anomalous points, respectively.

\subsection{3D convolutional (orthogonal) neural networks} \label{sub_sec:cnn_results}

Next, we test our primary prosed model - 3D convolutional neural networks with and without orthogonal components as the encoder and decoder in the autoencoder pipeline.  This is also the example we implement on the quantum hardware in Section~\ref{sec:quantum_hardware_exps}. We denote the fully classical 3D convolutional model as `3D-CNN' and the version containing (quantum) \textsf{OrthoConv3D} layers as `3D-QCNNs'. We deploy the same learning techniques here as for the FNNs in Section~\ref{sub_sec:fnn_results}, namely comparing point-estimate gradient descent (PE) with Bayesian learning, along with Monte Carlo Dropout (MCD) and Ensembling methods. Again, we begin by comparing the Bayesian learning approach for the classical and quantum(-inspired) 3D convolutional architecture in Table~\ref{tab:cnn_results_bayesian}.

We also compare the feedforward architectures from Section~\ref{sub_sec:fnn_results} with the 3D convolutional architectures in Figure~\ref{fig:ece_evaluation_bayesian_vs_non_bayesian}. We observe that the data-native architecture, including the 3D (classical) convolutional layer, outperforms the simpler feedforward networks, with and without (quantum) orthogonal layers. However, this could be due to the specific architecture we chose and may not be the conclusion more generally. On the other hand, the orthogonal parameterisation for the 3D-QCNN contains significantly fewer parameters than the 3D-CNN.

\section{Quantum hardware experiments} \label{sec:quantum_hardware_exps}

\begin{figure*}[htbp]
  \centering
  % Subfigure for Dataset 1
  \begin{subfigure}[t]{0.28\textwidth}
    \centering
    \begin{tikzpicture}
      \begin{axis}[
          xlabel={\sffamily Percentage (\%)},
          ylabel={\sffamily \(\text{MSE}(\widetilde{\boldsymbol{y}}, \boldsymbol{y})\)},
          grid=major,
          xmin=0, xmax=100,
        ymin=0, ymax=1,
          width=5.2cm,
          height=4.5cm,
      legend pos=north west, 
          legend style={font=\small}
      ]
        % Data points with error bars
        \addplot[
          color=gnomeblue,
            mark=square*, 
            mark options={line width=3, mark size=1},  
            error bars/.cd,
             y dir=both, y explicit,
        ]
        coordinates {
          (0, 0.0)         +- (0, 0)
          (10, 0.09379574)  +- (0, 0.021023883)
          (20, 0.17213114)  +- (0, 0.081777)
          (30, 0.2570897)   +- (0, 0.049347185)
          (40, 0.35746914)  +- (0, 0.033226706)
          (50, 0.45024833)  +- (0, 0.042637765)
          (60, 0.5021489)   +- (0, 0.046058603)
          (70, 0.63365155)  +- (0, 0.06957167)
          (80, 0.7374226)   +- (0, 0.051480412)
          (90, 0.78825015)  +- (0, 0.03150678)
          (100, 0.8829042)   +- (0, 0)
        };
        % Connecting line
        \addplot[
          color=gnomeblue,
          mark=,
          thick,
        ]
        coordinates {
          (0, 0.0)
          (10, 0.09379574)
          (20, 0.17213114)
          (30, 0.2570897)
          (40, 0.35746914)
          (50, 0.45024833)
          (60, 0.5021489)
          (70, 0.63365155)
          (80, 0.7374226)
          (90, 0.78825015)
          (100, 0.8829042)
        };
        \addlegendentry{\textsf{OrthoLinear}}
      \end{axis}
    \end{tikzpicture}
    \caption{}
  \end{subfigure}
  \qquad 
    % \hfill
  % Subfigure for Dataset 2
  \begin{subfigure}[t]{0.28\textwidth}
    \centering
    \begin{tikzpicture}
    \centering
      \begin{axis}[
          xlabel={\sffamily Percentage (\%)},
          ylabel={\sffamily \(\text{MSE}(\widetilde{\boldsymbol{\mathcal{Y}}}, \boldsymbol{\mathcal{Y}})\)},
          grid=major,
          xmin=0, xmax=100,
          ymin=0, ymax=4e-3,
          width=5.2cm,
          height=4.5cm,
              mark=o*, 
            mark options={line width=3, mark size=1},
          legend pos=north west, 
          legend style={font=\small}
      ]
        \addplot[
          color=gnomeorange,
          mark=*,
          error bars/.cd,
             y dir=both, y explicit,
        ]
        coordinates {
          (0, 0.0)             +- (0, 0)
          (10, 0.00030940142)   +- (0, 6.352416e-05)
          (20, 0.00056176225)   +- (0, 8.79111e-05)
          (30, 0.00092848635)   +- (0, 0.00014135808)
          (40, 0.0014027094)    +- (0, 0.00020038123)
          (50, 0.0015578003)    +- (0, 0.00023935977)
          (60, 0.002112804)     +- (0, 0.00012301456)
          (70, 0.0024042246)    +- (0, 5.2160307e-05)
          (80, 0.0026879406)    +- (0, 4.2984924e-05)
          (90, 0.003087505)     +- (0, 9.043532e-05)
          (100, 0.0034488444)   +- (0, 0)
        };
        \addlegendentry{\textsf{OrthoConv3D}}
        \addplot[
          color=gnomeorange,
          mark=none,
          thick,
        ]
        coordinates {
          (0, 0.0)
          (10, 0.00030940142)
          (20, 0.00056176225)
          (30, 0.00092848635)
          (40, 0.0014027094)
          (50, 0.0015578003)
          (60, 0.002112804)
          (70, 0.0024042246)
          (80, 0.0026879406)
          (90, 0.003087505)
          (100, 0.0034488444)
        };
      \end{axis}
    \end{tikzpicture}
    \caption{}
  \end{subfigure}
  % \hfill
   \qquad
  % Subfigure for Dataset 3
  \begin{subfigure}[t]{0.28\textwidth}
    \centering
    \begin{tikzpicture}
      \begin{axis}[
          xlabel={\sffamily Percentage (\%)},
          ylabel={\sffamily \(\text{MSE}(\widetilde{\boldsymbol{\mathcal{Z}}}, \boldsymbol{\mathcal{Z}})\)}, 
          grid=major,
            xmin=0, xmax=100,
          ymin=0, ymax=1.5e-8,
          width=5.2cm,
            height=4.5cm,
          legend pos=north west, 
          legend style={font=\small}
      ]
        \addplot[
          color=gnomegreen,
            mark=x, 
            mark options={line width=3, mark size=3},
          error bars/.cd,
             y dir=both, y explicit,
        ]
        coordinates {
          (0, 0.0)              +- (0, 0)
          (10, 8.2225443e-10)    +- (0, 1.3939474e-09)
          (20, 1.9549753e-09)    +- (0, 1.5525748e-09)
          (30, 4.295461e-09)     +- (0, 3.6427064e-09)
          (40, 2.945828e-09)     +- (0, 2.1609399e-09)
          (50, 4.607139e-09)     +- (0, 2.579102e-09)
          (60, 8.98478e-09)      +- (0, 4.3884585e-09)
          (70, 4.5720854e-09)    +- (0, 2.8485172e-09)
          (80, 6.3103265e-09)    +- (0, 1.0060328e-09)
          (90, 7.784574e-09)     +- (0, 4.290225e-09)
          (100, 9.85983e-09)     +- (0, 0)
        };
        \addlegendentry{\textsf{Q. Autoencoder}}
          color=gnomegreen,
          mark=none,
          thick,
        ]
        coordinates {
          (0, 0.0)
          (10, 8.2225443e-10)
          (20, 1.9549753e-09)
          (30, 4.295461e-09)
          (40, 2.945828e-09)
          (50, 4.607139e-09)
          (60, 8.98478e-09)
          (70, 4.5720854e-09)
          (80, 6.3103265e-09)
          (90, 7.784574e-09)
          (100, 9.85983e-09)
        };
      \end{axis}
    \end{tikzpicture}
    \caption{}
  \end{subfigure}
  \caption{\textsf{
  \textbf{Mean squared error (MSE) for different model ingredients.} Take \(\boldsymbol{\mathcal{X}}\) to be the full voxel, with reference slice \(\mathbf{X}_{j, k} := [\boldsymbol{\mathcal{X}}]_{j, k, 14}\). The input to any given orthogonal quantum layer within the architecture is a vector, denoted \(\boldsymbol{x}\), created as an \(8\) dimensional feature vector from the 3D kernel. \textbf{(a)} \(\boldsymbol{y}\) is the output directly from an orthogonal quantum layer, \(\boldsymbol{y} = \text{OrthoQNN}(\boldsymbol{x})\).  \textbf{(b)} \(\boldsymbol{y}\) is the output from the full 3D convolutional layer, \(\boldsymbol{\mathcal{Y}} = \text{OrthoConv3D}(\boldsymbol{\mathcal{X}}')\), where \(\boldsymbol{\mathcal{Y}}, \boldsymbol{\mathcal{X}}'\) are intermediate voxels in the pipeline. \textbf{(c)} \(\boldsymbol{y}\) is the output from the full autoencoder model with quantum orthogonal ingredients, \(\boldsymbol{\mathcal{Z}} = \text{QAE}(\boldsymbol{\mathcal{X}})\) where \(\boldsymbol{\mathcal{Z}}\) is ideally a reconstructed version of \(\boldsymbol{\mathcal{X}}\). The x-axes show the \(\%\) of circuits (out of \(256\)) which are included in the model, that ran on IBM Brisbane, the rest being simulated exactly. As the relative weight of the hardware results decreases within each respective pipeline, it has less influence on the error.  
  }
  }
  \label{fig:hardware_exp_2_full_pipeline}
\end{figure*}

\subsection{Computing fidelities} \label{sec:quantum_hardware_exps_fid}
To begin, we run a simple experiment to test the effects of 1) measurement noise, and 2) the data when running a part of the hybrid architecture on a quantum computer. The results of this are shown in Fig.~\ref{fig:hardware_fidelity_experiment}. Specifically, we use a simple orthogonal quantum layer, \(\textsf{OrthoQNN}\) on input vectors, \(\{\boldsymbol{x}_i\}_{i=1}^{8}\) to produce outputs, \(\boldsymbol{y}_i = \textsf{OrthoQNN}(\boldsymbol{x}_i), \forall i\). We use a data loader to create the initial state (specifically the `parallel' loader), \(\ket{\boldsymbol{x}_i}\), operate with an orthogonal layer with fixed angles (using the `pyramid' layout), \(U_O(\boldsymbol{\theta}^*)\) and measure to extract the output, \(\widetilde{\boldsymbol{y}}_i\), and compare with the ideal output state, \(\ket{\boldsymbol{y}}\) via the fidelity, \(F(\rho_{\boldsymbol{y}}, \ket{\boldsymbol{y}}\bra{\boldsymbol{y}}) := \bra{\boldsymbol{y}}\rho_{\boldsymbol{y}}\ket{\boldsymbol{y}}\) where \(\rho_{\boldsymbol{y}}\) is the noisy state produced by the device according to \(\boldsymbol{y}\). To approximate this, we compute \(\widetilde{F}(\rho_{\boldsymbol{y}}, \ket{\boldsymbol{y}}\bra{\boldsymbol{y}}) \approx (\widetilde{\boldsymbol{y}}\cdot \boldsymbol{y})^2, \widetilde{\boldsymbol{y}} := \sqrt{\widetilde{\boldsymbol{p}}}\), where \(\widetilde{\boldsymbol{p}}\) is the post-processed probability distribution over unary bitstrings seen by \(M\) measurement shots. Unary post-selection acts as an algorithmic method of error-mitigation (EM) to complement the inbuilt methods of compilation optimisation and error-mitigation provided by IBM. We use the default EM level, which is readout EM only, and the highest level of compiling optimisation. The compiler optimisation is needed primarily for SWAP routing in the data-loading state, since the parallel loader has non-nearest neighbour connectivities.

We perform two sets of experiments. The first, seen in Fig.~\ref{fig:hardware_fidelity_experiment} (top row) chooses vectors, \(\{\boldsymbol{x}_i\}_{i=1}^{8}\) uniformly at random to pass through the ortholinear layers. The second set Fig.~\ref{fig:hardware_fidelity_experiment} (bottom row) chooses a slice from the anomalous object data and downscales to \(8 \times 8\). Each vector, \(\{\boldsymbol{x}_i\}_{i=1}^{8}\), then corresponds to a single row or column from the data. We run both sets of experiments with either \(M = 1k\) or \(M=10k\) measurement shots.

We can make two observations from the Figure. Firstly, in the absence of structure coming from the data, increasing the measurement shots does not tend to help, which with real data it does. One possible explanation is that the real data presents features that increased resolution via more measurements can help resolve. These features are more likely to dominate the fidelity, particularly for small-scale experiments. The second observation is the higher average fidelity overall between the \(M=10k\) experiments with real data. The structure can perhaps be useful to inherently mitigate quantum noise by affecting gates which do not, on average, tend to affect the result strongly, both in the data-loading phase and the evolution phase. However, we cannot guarantee that these results are scalable to larger problem sizes.

\subsection{Robustness of anomaly detection pipeline with quantum hardware} \label{sec:experiment_relative_effect_anomaly_detection}

Next, we incorporate the quantum hardware component into the full anomaly detection pipeline. Here, we compare the output of various components of the quantum 3D CNN as progressively more ingredients are executed directly on the IBM quantum computer. In this set of experiments, we take an anomalous voxel and downscale to \(16 \times 16 \times 16\). This corresponds to \(16\) slices of \(16 \times 16\) pixels, as seen in Fig.~\ref{fig:flattened_anomaly_voxels}. We choose a slice which contains an anomaly (slice \(14\) in the Figure) and focus on components of the quantum AD pipeline which interact with this slice. 

The results for the three scenarios we study are seen in Fig.~\ref{fig:hardware_exp_2_full_pipeline}. 1) Orthogonal quantum layers acting directly on the voxel slice. This is the innermost part of the pipeline, which contains the quantum circuits (Fig.~\ref{fig:hardware_exp_2_full_pipeline}a). 2) A full orthogonal 3D convolutional block (Fig.~\ref{fig:hardware_exp_2_full_pipeline}b). 3) The complete autoencoder pipeline, including voxel pre- and post-processing layers (Fig.~\ref{fig:hardware_exp_2_full_pipeline}c).

In each case, we compare the output of the layer \emph{including} the hardware component (denoted with a tilde), to the true, noiseless, output computed via simulation. We plot the mean square error (MSE) in each case. For the output of the ortholinear layer (Fig.~\ref{fig:hardware_exp_2_full_pipeline}a), the output of each circuit is a vector, \(\widetilde{\boldsymbol{y}}, \boldsymbol{y}\) respectively. Within the 3D convolutional pipeline, we use a \(2\times 2 \times 2\) 3D kernel with stride \(1\). In the downscaled version, these parameters correspond to a circuit per pixel in the slice (\(14\)), which is \(16 \times 16 = 256\), on \(2\times 2 \times 2 = 8\) qubits. 

To generate the results in the Figure, we successively increase the fraction of these circuits (\(\%\)) which are executed on the quantum hardware. As can be seen from Fig.~\ref{fig:hardware_exp_2_full_pipeline}a), the closeness of the hardware results to their true values decreases linearly as the fraction of hardware-ran circuits increases. 

We repeat similar experiments, but now compare the effect of the quantum component on increasingly larger parts of the full pipeline, with the OrthoConv3D block in Fig.~\ref{fig:hardware_exp_2_full_pipeline}b) and the full quantum autoencoder output in Fig.~\ref{fig:hardware_exp_2_full_pipeline}c). In these latter cases, we measure the MSE between the voxel output components, \(\boldsymbol{\mathcal{Y}}\) and \(\boldsymbol{\mathcal{Z}}\) respectively. We see that as the quantum hardware components reduce their relative influences on the model component in question, the deviation of the overall output from the true values significantly decreases. In the case of the OrthoConv3D layer, the maximum MSE reaches only \(\sim 3.5\times 10^{-3}\), while for the full pipeline the overall deviation in MSE of the reconstructed object is only \(\sim 1\times 10^{-8}\) - even when all \(256\) component circuits are executed on the hardware. For all of these experiments, we use \(M=5k\) shots, and \(8\) qubits from the \(127\) qubit IBM Brisbane quantum computer.

\section{Conclusion and Outlook}
\label{Conclusion}
Traditional methods for detecting defects in additive manufacturing are time-consuming and labour-intensive. Machine learning and artificial intelligence techniques offer an alternative solution to this problem by quickly identifying defects in an easily adaptable manner. Furthermore, quantum and quantum-inspired computing have the potential to uplift the expressivity and efficiency of such ML methods via quantum machine learning. In this work, we have demonstrated the feasibility of this for our particular use case. By combining Bayesian learning with orthogonal neural operations in a three-dimensional architecture, defects in additive manufactured parts can be detected successfully, along with measuring the uncertainty of those detections. Furthermore, the quantum nativity of our architectures enables smooth deployment of such methods, once suitably sized and quality quantum processing units are developed. In conclusion, this work adds another vital element to the growing field of anomaly detection using quantum computers for real-world use cases.

\bibliographystyle{alpha}
\bibliography{sample}

\end{document}